\documentclass[journal,onecolumn]{IEEEtran}

\usepackage{graphicx}
\usepackage{changepage}
\usepackage{romannum}
\usepackage{caption}
\usepackage{empheq}
\usepackage{amsmath}
\usepackage{mathtools,cuted}
\usepackage[ruled,vlined]{algorithm2e}
\usepackage[noend]{algpseudocode}

\usepackage{siunitx}
\usepackage{amssymb}
\usepackage{booktabs}
\usepackage{xcolor}
\newcommand{\etal}{\textit{et al}.}

\title{Spatio-Temporal Bayesian Learning for Mobile Edge Computing Resource Planning in Smart Cities}

\author{Laha Ale, Ning Zhang, Scott A. King, Jose Guardiola}

\begin{document}
\maketitle

\begin{abstract}
A smart city improves operational efficiency and comfort of living by harnessing techniques such as the Internet of Things (IoT) to collect and process data for decision making. To better support smart cities, data collected by IoT should be stored and processed appropriately. However, IoT devices are often task-specialized and resource-constrained, {and thus,} they heavily rely on online resources in terms of computing and storage to accomplish various tasks. { Moreover, these cloud-based solutions often centralize the resources and are far away from the end IoTs and cannot respond to users in time due to network congestion when massive numbers of tasks offload through the core network. Therefore, by} decentralizing resources spatially close to IoT devices, mobile edge computing (MEC) can reduce latency and improve service quality for a smart city, where service requests can be fulfilled in proximity. As the service demands exhibit spatial-temporal features, deploying MEC servers at optimal locations and allocating MEC resources play an essential role in efficiently meeting service requirements in a smart city. In this regard, it is essential to learn the distribution of resource demands in time and space. In this work, we { first} propose a spatio-temporal Bayesian hierarchical learning approach to learn and predict the distribution of MEC resource demand over space and time to facilitate MEC deployment and resource management. { Second}, the proposed model is trained and tested on real-world data, and the results demonstrate that the proposed method can achieve very high accuracy. { Third}, we demonstrate an application of the proposed method by simulating task offloading. { Finally, the simulated results show that resources allocated based upon our models' predictions are exploited more efficiently than the resources are equally divided into all servers in unobserved areas.}

\end{abstract}

\IEEEpeerreviewmaketitle

\section{Introduction}
A smart city holds the promise to improve living comfort and operational efficiency in urban areas, which incorporates massive Internet of Things (IoT) for collecting data to manage assets and provide various services efficiently~\cite{smart_cit, Puliafito2019}. With ubiquitous IoT devices, various types of data can be collected and processed for making better decisions. Moreover, the majority of IoT devices are equipped with minimal resources and rely heavily on remote servers to process their tasks by offloading content and computation through the core network. { Ideally, the IoTs can rely on the cloud-based servers and process various offloaded tasks with efficient cloud resource planing~\cite{Fei2020} or cloud-assisted heterogeneous networks~\cite{Zhang7120046}. However, cloud-based solutions often centralize the resources and are far away from the end IoTs and cannot respond to users in time due to many challenges. First, the network will be congested with the increase in the number of IoTs and offloaded tasks through core networks. Consequently, the core networks are inevitably overloaded and cause congestion and increase the remote server response latency. Moreover, the number of IoTs is growing exponentially, and many IoTs are sensitive to the response time.} 

To mitigate the above issues, Mobile Edge Computing (MEC)~\cite{mec2018} is proposed to process tasks in proximity by distributing resources to local edge servers, which are spatially adjacent to IoT devices. In other words, the local MEC servers address most of the offloaded tasks from nearby IoT devices and save the cost to transmit all the received tasks to remote servers~\cite{task_offloading}; therefore, MEC can considerably reduce the burden of core networks. However, the MEC servers are required to intelligently exploit the {limited computation power} and storage capacity they have~\cite{challenges}. Therefore, knowing the storage and computing resource demands from IoT devices is essential not only to finding optimal locations to place MEC servers but also to dynamically and efficiently allocate resources. Hence, modeling for IoT devices and predicting the space-time distribution of task offloading is crucial for deploying and exploiting MEC resources efficiently. Surprisingly, despite some works regarding the IoT data collection and analysis, only scarce research exists regarding modeling and predicting the space-time distribution of MEC resources.

As mentioned above, geospatial data science methods are applied to IoT data collection and analysis, as well as its applications. Kamilari \etal~\cite{Kamilaris2018} give a geospatial analysis of IoT applications, including transportation, agriculture, and sport and geospatial analytical methods, including geometric measures, analytical operations, and network analysis. A fire evacuation application in buildings ~\cite{Liu2014} uses IoT sensors to monitor and collect data, and a Geographic Information System (GIS) to identify buildings and locations inside those buildings. Additionally, methods to address high-speed streaming data from IoT devices have been surveyed by Zhang \etal~\cite{Armstrong2019}; they summarized the challenges and pointed out possible directions to address those issues. However, they did not provide details of the possible solutions regarding the mentioned challenges. { Efficient data share frameworks~\cite{Yuan2018,Luo2019} for Vehicular Networks can reduce the burden of the core network in smart cities. However, users may have concerns related to sharing privacy-sensitive data through MEC. Also, it is challenging to deploy a framework that allows other types of IoTs to share data or computational resources because the IoTs are resource limited.}

To support various spatially distributed IoT devices, limited MEC methods have been proposed to mitigate the challenges. To improve geospatial data collection efficiency, Cao \etal~\cite{Cao2019} proposed a method to reduce data collection effort by replacing GPS data with MEC server data. In other words, MEC server locations have represented clusters of IoT locations, which may compromise the spatial resolution. {Zhang et.al~\cite{DBLPZhang92018} proposed a Weighted Expectation Maximization (WEM) model to deploy UAV-based MEC servers efficiently. However, they do not leverage the spatial correlation as an essential feature in our work. }Yan \etal~\cite{Guo2019} have studied a MEC server placement suitability problem to find the optimal locations for the MEC server based on the distribution of the workload (i.e., the density of requests from IoTs).  Similarly, a method for MEC server placement optimization has been presented by Tran \etal~\cite{Tran2018}, and plausible locations are derived by minimizing communication costs. Additionally, geospatial resource coordination for MEC ~\cite{Casadei2019} and multi-armed bandit~\cite{Chen2018} are adopted to solve spatial-temporal MEC placement queries for MEC services. Overall, those methods are based on the geo-clustering of IoTs~\cite{Bouet2018} to find the optimal placement for MEC servers and resources. However,  the distribution of offloading tasks from the IoTs varies significantly over the { space and time, and methods based on clustering spatially only may not be efficient for MEC resource provision.}

Although methods for spatio-temporal modeling and predicting specialized for MEC are insufficient, various general-purpose statistics and learning methods are proposed to solve comparable problems. Therefore, it would not be difficult to change the contours of that map of the MEC by adapting methods from geospatial statistics~\cite{Banerjee2015} and artificial intelligence, including machine learning and {Deep Learning (DL)}~\cite{Lecun2015}. Statistics models, especially Bayesian clan models, are adopted for modeling and predicting spatio-temporal data~\cite{Smith2003}. Note that statistical models do not demand a large number of historical data records to fit the models; however, they cannot efficiently leverage a massive amount of data and capture complex patterns of the real-world MEC applications. 

Therefore, Deep Learning (DL) methods are adapted to learn and predict spatio-temporal data. The DL methods can extract robust patterns from complex and noisy data from the wireless network~\cite{Wang2017}. Additionally, DL models can process a vast amount of data by leveraging multiple nodes computing from High-Performance Computing (HPC) and parallel computing from Graphics Processing Units (GPU). However, the MEC servers are typically equipped with limited resources that cannot support heavy models like DL types. Besides, DL models require human labor efforts to collect and label massive training data from DRL models, which increases the cost of using the DL models. Furthermore, DL models are extremely challenging to interpret the relation of considered factors and predicted results.

Few real-world datasets exist that provide testing beds for researchers to verify DL and other models. Fortunately, we have access to two datasets, including the Shanghai telecom dataset published by Yan \etal~\cite{Guo2019} and multi-source dataset of urban life in the city of Milan and the province of Trentino~\cite{Barlacchi2015}. The dataset from Shanghai telecom is clean data, including six months with millions of user connection records but it is {low-dimensional.} The second dataset from Italy has relatively more dimensions for mining and analysis from the geospatial MEC perspective.

To predict spatio-temporal distributions of resource demands and support MEC network resource deployment in smart cities, we adopt a hierarchical spatio-temporal Bayesian method~\cite{Bakar2015} to model and predict the space-time distributions of requests. The adopted model can effectively process millions of records without costing computational resources beyond those MEC servers could support. Additionally, the model is designed online-learning fashion; it keeps learning and improving its performance over time. Furthermore, the hierarchical model can capture moderately intricate patterns of data and produce stable and accurate predictions. The prediction results can be used to find optimal { locations for the MEC servers and for online resources provision dynamic} support smart city operation. \emph{\bf The main contributions of this work are:}

\begin{itemize}
     \item A data clustering method is developed to aggregate and visualize the space-time distribution of resource demands on a network.
    \item A hierarchical spatio-temporal Bayesian method is designed to model and predict the spatial-temporal distributions of MEC workload patterns.
    \item The proposed model is tested on a real-world dataset from the China Telecom company in Shanghai. The data includes over four million records of six months of user connections to towers in the Shanghai area.
  
    \item We create a simulation of IoT devices and their demands in unobserved areas, based upon historical data, and show how our predictive model aids placement of mobile MEC servers to utilize the limited resources more efficiently.
    
\end{itemize}

The remainder of the paper is organized as follows. Section~\ref{sec:2} presents the system model and problem formulation; the adopted method with hierarchical Bayesian models are introduced in Section~\ref{sec:3} ; results and analysis are provided in ~\ref{sec:4}; in Section~\ref{sec:5}, a simulation demonstrates a possible application of the adopted method, and the future work has been discussed in Section~\ref{sec:6}; and Section~\ref{sec:7} provides conclusions.

\section{System Model and Problem Formulation}
\label{sec:2}

\begin{figure} [!h]
    \centering
    \includegraphics[width=5in]{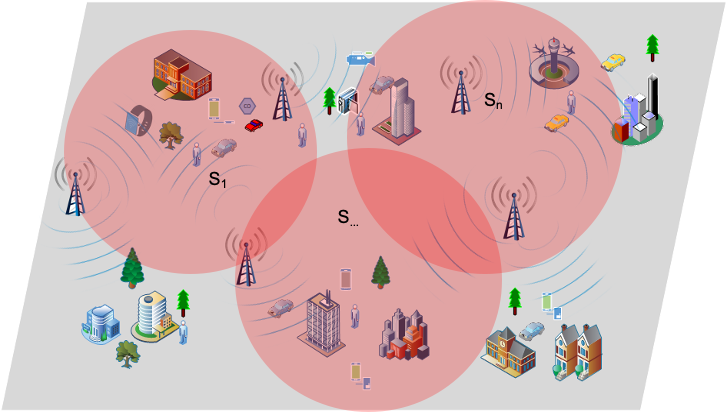}
    \caption{System Model.}
    \label{fig:problem}
\end{figure}

In this section, we present the descriptions of the system model and problem formulation. As shown in Fig.\ref{fig:problem}, the IoT devices are spatially distributed and connected to the nearby Base Stations (BS), which are the MEC server hosts. Additionally, the IoT devices are moving around along with their hosts, and their resource demands change over time. Therefore, the distributions of the resource demands of IoT devices are typically unknown to the network providers. To simplify the descriptions of modeling and predicting, we divide continuous-time into discrete time slots. 
The time slots are denoted as $\mathcal{T} = \{1,\dots, t,\dots, \mathcal{A}\}$. Similarly, we divide the space into small unit {areas denoted} as $S = \{s_1, s_2, \dots, s_n\}$. Note that each unit area could contain multiple towers. Further, an observation and corresponding value of site $s_i$ at time slot $t$ are denoted by $o(s_i,t)$ and  $v(s_i,t)$, respectively. The observations are the input parameters, including the number of users, MEC servers, other related feature values, and the corresponding current value, which is the workload in the unit area. Although the corresponding value is the workload defined by the ratio of the sum of the connections to the number of towers in this work, it could be generalized to many other MEC value predictions, including the number of IoT devices, tasks, the total required CPU cycles, and the size of total downloaded files. 
Further, the total observation set and corresponding value set can be denoted as { $O_t = (o(s_1,t),\dots,o(s_n,t))^{T}$ and $V_t = (v(s_1,t),\dots,v(s_n,t))^{T}$, respectively, and $(\cdot)^{T}$ is the transpose function.} Let $o^*$ denote missing observations, and $\mathcal{D}=n \mathcal{A}$ be the total number of historical observation records.

In geospatial modeling, we need to consider not only non-spatial effects but also spatial variances. The nugget effect, refers to non-spatial variance, and is given by $\boldsymbol{\epsilon}_t = (\epsilon(s_1,t),\dots,\epsilon(s_n,t))^{T}$ for all the sites at time $t$. We assume the nugget effects for the sites are independent and have a normal distribution $\mathcal{N}(0,\sigma_\epsilon^2I_n)$, where $\sigma_\epsilon^2$ is the known variance and $I_n$ is the $n\times n$ identity matrix. Additionally, the spatial-temporal random effects are denoted by $\boldsymbol{\eta}_t=(\eta(s_1, t), \ldots, \eta(s_n, t))^T$; these random effects are assumed to follow $\mathcal{N}\left(\mathbf{0}, \Sigma_{\eta}\right)$ independently in time, where $\Sigma_{\eta}=\sigma_\eta^2 \zeta_\eta$,  $\sigma_\eta^2$ is the site constant spatial variance, and $\zeta_\eta$ is the spatial correlation matrix obtained from the general Matern correlation function given by:

\begin{equation}
\label{eqn_matern}
\begin{split}
\kappa\left(\mathbf{s}_{i}, \mathbf{s}_{j} ; \phi, \nu\right)=\frac{1}{2^{\nu-1} \Gamma(\nu)}\left(2 \sqrt{\nu}\left\|\mathbf{s}_{i}-\mathbf{s}_{j}\right\| \phi\right)^{\nu} K_{\nu}\left(2 \sqrt{\nu}\left\|\mathbf{s}_{i}-\mathbf{s}_{j}\right\| \phi\right), \quad \phi>0, \nu>0
\end{split}
\end{equation}
where, $s_i$ and $s_j$ are the locations in the correlation matrix indexed by $i$ and $j$; $\Gamma(\nu)$ is a regular Gamma function~\cite{Sebah2002}; $K_{\nu}$ is a modified Bessel function of the second kind~\cite{Baricz2010} with order $\nu$, where $\phi$ controls the spatial correlation decay, and $\nu$ controls the smoothness of the random field. Intuitively, $K_{\nu}$ controls correlation of $s_i$ and $s_j$ decreasing as the $2\sqrt{\nu}\left\|\mathbf{s}_{i}-\mathbf{s}_{j}\right\| \phi$ increases, and $\nu$ controls the decrease speed in function $K_{\nu}$, where $\left\|\mathbf{s}_{i}-\mathbf{s}_{j}\right\|$ is the distance between $s_i$ and $s_j$.  According to the First Law of Geography, the closer two observations, the more related. The spatial correlation decay parameter $\phi$, is the effective range parameter that controls the degree to which spatial correlation decreases with an increase in distance. 

Considering the above system model description, the problem formulation is very straightforward. Specifically, the target problem is to predict the distribution of MEC workload over space-time. Let $m$ be the number of covariates, and the input features including the intercepts can be denoted as  $\mathbf{X}_t=\{x_0,x_1,\dots,x_{m-1}\}$. In practice, we can add a column to the feature matrix, and all the values in that column can be set to 1 as the placeholder for the intercept. Generally, the features contain more than one factor that may affect the reference outputs. We extract several features, including the number of unique users, base stations, and total connection time. Further, let $\beta=(\beta_0,\beta_1,\dots,\beta_{m-1})$ be an $m\times 1$ vector { containing} regression coefficients corresponding to the feature $\mathbf{X}_t$. To model and predict the distributions of the workload over space and time, the model is required to predict values that include the time series values $V_{t'} = (v(s_1,t'),\dots,v(s_n,t'))^{T}$ of the observed and unobserved areas $S' = \{s_1', s_2', \dots, s_n'\}$. Therefore, we need to set up a model that will be able to predict both spatial and temporal values based on the observed features and current 
corresponding values. 

Similar to linear regression in standard machine learning, the relation of $\mathbf{O}_t$ and $\mathbf{V}_t$ can be modeled as a hierarchical Auto-Regressive (AR) model given by:
\begin{equation}
\label{eqn_ar1}
\begin{cases}
\mathbf{O}_t=\mathbf{V}_t+\boldsymbol{\epsilon}_t, \\
\mathbf{V}_t= \rho \mathbf{V}_{t-1} + \mathbf{X}_t \boldsymbol{\beta}+\boldsymbol{\eta}_t 
\end{cases}
\end{equation}
Or, by a hierarchical Gaussian Process (GP) model given by: 
\begin{equation}
\label{eqn_gp1}
\begin{cases}
\mathbf{O}_t=\mathbf{V}_t+\boldsymbol{\epsilon}_t, \\
\mathbf{V}_t= \mathbf{X}_t \boldsymbol{\beta}+\boldsymbol{\eta}_t 
\end{cases}
\end{equation}
We assume that each time slot, $t\in \mathcal{T}$ has a nugget effect $\epsilon_t$ and spatial-temporal effect,  $\eta_t$. Let $\pi(\boldsymbol{\theta})$ denote the prior distribution (presented in the following section), and let $\rho \in (-1,1)$ be the unknown temporal correlation. 

To predict the distributions of resources demands, the proposed method aims to minimize the errors measured by the sum of the square or absolute values of differences between predicted values and observed values. We have a testing dataset with $d$ records indexed by $i$, the predicted value for the $i^{th}$ record is $\hat{v}_i = \rho v_{i-1} + X_i \beta+\eta_i$, and the corresponding true value is $v_i$. Specifically, the goal of the proposed model is to minimize a loss function of the errors, which can be measured by several metrics, including Mean Squared Error (MSE), Mean Absolute Error (MAE), Mean Absolute Percentage Error (MAPE), and Relative Mean Separation (RMSEP) described as follows: 
\begin{equation}
\label{eqn_errors}
\begin{split}
MSE = \frac{1}{d}\sum_{i=1}^d(\hat{v}_i - v_i)^2, \ 
MAE = \frac{1}{d}\sum_{i=1}^d|\hat{v}_i - v_i|, \\
MAPE = \frac{1}{d}\sum_{i=1}^d|\frac{\hat{v}_i - v_i}{v_i}| ,\ 
RMSEP = \frac{\sum_i^d(\hat{v}_i - v_i)^2}{\sum_{i=1}^d(\Bar{v}_p - v_i)^2} 
\end{split}
\end{equation}
where $\Bar{v}_p$ is the mean of the predicted values.

Note that these metrics are also known as the loss functions in standard machine learning. Although we can use the above functions to evaluate the performance of the model, the Predictive Model Choice Criteria (PMCC) function is adopted to loss function in the training process. The PMCC computes the loss function using errors and variances of the distance between predicted values and observed values, which provide more information for training the model. The details of the PMCC are presented in the next section. 

\section{Method}
\label{sec:3}
In this section, we introduce the method for training the Bayesian Auto-Regressive (AR) random-process model. To reduce the training effort and minimize the prediction errors, we need to provide suitable prior distributions and a training and predicting process.

\subsection{Spatio-Temporal Bayesian Modeling}

As defined in the previous section, the AR model requires initial values for each site before we start the training process. We specify the initial values using a mean $\boldsymbol{\mu}$ and covariance $\sigma^2 \zeta_0$, where the correlation matrix $\zeta_0$ can be obtained by the Matern correlation function (Eq.\ref{eqn_matern}). For the sake of simplicity, let $\boldsymbol{\theta}=\left(\boldsymbol{\beta},\rho, \sigma_{\epsilon}^2, \sigma_{\eta}^2, \phi, \nu, \boldsymbol{\mu}, \sigma^2 \right)$ denote all the parameters of the model and $\pi(\boldsymbol{\theta})$ denotes the prior distribution. Furthermore, the logarithm of the joint posterior distribution of the parameters and the missing data for the AR model can be given by:

\begin{equation}
\label{eqn_ar}
\begin{split}
\log \pi\left(\boldsymbol{\theta}, \mathbf{V}, \mathbf{o}^* | \mathbf{o}\right) \propto & -\frac{N}{2} \log \sigma_{\epsilon}^2-\frac{1}{2 \sigma_\epsilon^2}  \sum_{t=1}^{T}\left(\mathbf{O}_{t}-\mathbf{V}_{t}\right)^{tr} \left(\mathbf{O}_{t}-\mathbf{V}_t\right)-\frac{ T}{2} \log \left|\sigma_{\eta}^2 \zeta_{\eta}\right| \\ &
-\frac{1}{2 \sigma_{\eta}^2}  \sum_{t=1}^{T}\left(\mathbf{V}_t-\rho \mathbf{V}_{t-1}-\mathbf{X}_t \boldsymbol{\beta}\right)^{tr} \zeta_\eta^{-1}  \left(\mathbf{V}_{t}-\rho \mathbf{V}_{t-1}-\mathbf{X}_t \boldsymbol{\beta}\right) 
-\frac{1}{2}  \log \left|\sigma^2 \zeta_{0}\right| \\ & -\frac{1}{2\sigma^2}\left(\mathbf{O}_0-\boldsymbol{\mu}\right)^{tr} \zeta_{0}^{-1} \left(\mathbf{O}_0-\boldsymbol{\mu}\right)+\log \pi(\boldsymbol{\theta})
\end{split}
\end{equation}
The $\pi(\boldsymbol{\theta})$ distributions are described in the following subsection, and the details for deriving the joint posterior distribution parameters are proved in Appendix A.

 The logarithm of the joint posterior distribution of the parameters and the missing data for the hierarchical GP model can be given by:
\begin{equation}
\label{eqn_gp}
\begin{split}
\log \pi\left(\boldsymbol{\theta}, \mathbf{V}, \mathbf{o}^* | \mathbf{o}\right) \propto & -\frac{N}{2} \log \sigma_{\epsilon}^2-\frac{1}{2 \sigma_\epsilon^2}  \sum_{t=1}^{T}\left(\mathbf{O}_{t}-\mathbf{V}_{t}\right)^{tr} \left(\mathbf{O}_{t}-\mathbf{V}_t\right)-\frac{ T}{2} \log \left|\sigma_{\eta}^2 \zeta_{\eta}\right| \\ &
-\frac{1}{2 \sigma_{\eta}^2}  \sum_{t=1}^{T}\left(\mathbf{V}_t-\mathbf{X}_t \boldsymbol{\beta}\right)^{tr} \zeta_\eta^{-1}  \left(\mathbf{V}_{t}-\mathbf{X}_t \boldsymbol{\beta}\right) 
-\frac{1}{2}  \log \left|\sigma^2 \zeta_{0}\right| +\log \pi(\boldsymbol{\theta})
\end{split}
\end{equation}

\subsection{Prior Distributions}

As with regular Bayesian models, the prior settings are considered important components of the adopted model. In the adopted hierarchical Bayesian model, it is necessary to provide the prior distributions and initialize them. In this work, the majority of the parameters are assumed normally distributed. Therefore, we can initialize the priors using independent distributions with their corresponding means and variances. For instance, the distribution of unknown temporal correlation, $\rho$ and the regression coefficient, $\beta$ can be initialized using means $(\mu_\rho,\mu_\beta)$ and variances $(\sigma_\rho^2, \sigma_\beta^2)$, and all the means and variances can be set to $0$ and $10^{10}$, respectively. The other parameters, including the nugget effect, $\sigma_\epsilon^2$, spatio-temporal effect, $\sigma_\eta^2$, temporal correlation, $\rho$, and the coefficients, $\beta$, the rate of the distance correlation decay, $\phi$, can be modeled as a Gamma distribution with mean $\mu=a/b$ and variance $a/b^2$, where $a$ and $b$ are initialized using of 2 and 1, respectively. Moreover, the smoothness control parameter, $\nu$, can be modeled as a uniform distribution using values of $[0,1.5]$ with an increased step of $0.05$.

\subsection{Model Training}
In this subsection, we introduce the general process of updating the parameters in the adopted model during the training process. The parameters $\phi$ and $\nu$ are updated using the Metropolis-Hastings algorithm, and the rest of the parameters can be obtained using Gibbs sampling from full conditional distributions. As it is a customary Bayesian model in the context of Markov chain Monte Carlo (MCMC), { the current sampled parameter is conditional on the values of the rest of the parameters derived in the previous iteration and the true values of the observed data. In other words, the current prediction is based on the previous prediction and current observed values.} When we are using conjugate prior distributions, {we assume that the likelihood and the priors are composed of normal distributions; therefore, the posteriors are expected to be normally distributed.} Missing observations are sampled from their conditional distributions following the same iterative process. Suppose we have $k$ parameters in the model that can be obtained using the Gibbs sampler, and the parameters can be denoted as $\boldsymbol{\theta} = (\theta_1,\dots, \theta_k)$; therefore, they can be derived with Algorithm.\ref{alg_gibbs} as follows:

\begin{algorithm}[!h]

\SetAlgoLined
\KwResult{Updated parameters}
$\boldsymbol{\beta} \leftarrow (\beta_0,\dots,\beta_m)$; // coefficients of the features\\
$\boldsymbol{\theta} \leftarrow (\beta,\sigma_\epsilon^2, \sigma_\eta^2,\rho)$; // the length of $\theta$ is $k$ \\
 \For{$t \leftarrow 1$ to $T$} {

   Draw $\theta_1^{(t)}$ from: $p\left(\theta_1 | \theta_{2}^{(t-1)}, \theta_{3}^{(t-1)}, \ldots,\theta_{k-1}^{(t-1)},\theta_k^{(t-1)}, \mathbf{y}\right)$ ;\\
   Draw $\theta_2^{(t)}$ from: $p\left(\theta_2 | \theta_{1}^{(t)}, \theta_{3}^{(t-1)}, \ldots,\theta_{k-1}^{(t-1)},\theta_k^{(t-1)}, \mathbf{y}\right)$ ;\\
   \vdots
   Draw $\theta_{k-1}^{(t)}$ from: $p\left(\theta_{k-1} | \theta_{1}^{(t)}, \theta_{2}^{(t)}, \ldots,\theta_{k-2}^{(t)},\theta_k^{(t-1)}, \mathbf{y}\right)$ ;\\
   Draw $\theta_k^{(t)}$ from: $p\left(\theta_k | \theta_{1}^{(t)}, \theta_{k}^{(t)}, \ldots,\theta_{k-2}^{(t)},\theta_{k-1}^{(t)}, \mathbf{y}\right)$ ;\\
   
 }
 \caption{Gibbs sampler}
 \label{alg_gibbs}
\end{algorithm}

Furthermore, the expected $\theta_i$ conditional on $\mathbf{y}$ can be computed as:
\begin{equation}
\label{eqn_gp1}
\widehat{E}\left(\theta_{i} | \mathbf{y}\right)=\frac{1}{T-t_{0}} \sum_{t=t_{0}+1}^{T} \theta_{i}^{(t)}
\end{equation}
where the time from $t=0$ to $t_0$ are the \textit{burn-in} (skip sampling) iterations, and the parallel computation version can be given by:
\begin{equation}
\label{eqn_gp1}
\widehat{E}\left(\theta_{i} | \mathbf{y}\right)=\frac{1}{l\left(T-t_{0}\right)} \sum_{j=1}^{l} \sum_{t=t_{0}+1}^{T} \theta_{i, j}^{(t)} 
\end{equation}
where $l$ is the number of parallel computations.

\begin{algorithm}[!h]
\caption{Metropolis-Hastings}
\label{alg_met}
\SetAlgoLined
\KwResult{Updated parameters: $\phi$ and $\nu$}
 \For{$t \in 1:T $} {
 Draw $\theta^*$ from $q\left(\cdot | \boldsymbol{\theta}^{(t-1)}\right)$; \\ 
 // Compute the ratio:\\
 $r = \frac{h\left(\boldsymbol{\theta}^{*}\right)} {h\left(\boldsymbol{\theta}^{(t-1)}\right)} = \exp \left[\log h\left(\boldsymbol{\theta}^{*}\right)-\log h\left(\boldsymbol{\theta}^{(t-1)}\right)\right]$; \\
 \If{ $r \geq 1$}{set $\theta^{(t)}=\theta^{*}$;\\}
 \Else{set $ \boldsymbol{\theta}^{(t)}=\left\{\begin{array}{l}{\boldsymbol{\theta}^{*} \text { with probability } r} \\ {\boldsymbol{\theta}^{(t-1)} \text { with probability } 1-r}\end{array}\right.$ \\
 }
 }
 
\end{algorithm}
As the spatial correlation decay parameter $\phi$ and random smoothness control parameter $\nu$ only accept positive values, we propose a Gamma prior distribution for $\phi$ and a uniform distribution for $\nu$. The prior distributions $p(\boldsymbol{\theta})$ and likelihood $f(\mathbf{y}|\boldsymbol{\theta})$ are not a conjugate pair and the full conditional distributions are not always available. Therefore, the Gibbs sampler method cannot be used for sampling these variables, both $\phi$ and $\nu$ should be computed using the Metropolis-Hastings algorithm (Alg.~\ref{alg_met}) as it only requires a function that is proportional to the distribution function of the sample, i.e $p(\boldsymbol{\theta}|\mathbf{y}) \propto h(\boldsymbol{\theta})=f(\mathbf{y}|\boldsymbol{\theta})p(\boldsymbol{\theta})$. The algorithm starts drawing samples from  a candidate density $q(\boldsymbol{\theta}^*|\boldsymbol{\theta}^{(t-1)})$ such that $q(\boldsymbol{\theta}^*|\boldsymbol{\theta}^{(t_1)}) = q(\boldsymbol{\theta}^{(t-1)}|\boldsymbol{\theta}^*)$. Furthermore, the parameters of the proposed model are sampled from a normal distribution with a mean $\mu_p$ and variance $\sigma_p^2$, which need to be adjusted to have an acceptance rate between 15\% and 40\% as suggested by Gelman \etal~\cite{Gelman2003}.

The above two algorithms describe the general ideas of the proposed learning processes. The details of these derivations of the parameters are shown in the Appendix A.
\subsection{Prediction and Evaluation}


As the model is a spatial-temporal Bayesian model, the prediction contains spatial and temporal parts. The spatial prediction is aimed to predict the MEC workload for the unobserved locations. Temporal prediction is used to predict future values at the selected locations, which can be observed or unobserved. Overall, the posterior predictive distribution generates marginal distributions for obtaining predicted values. The posterior predictive distribution can be given by:
\begin{equation}
\label{eqn_predict}
\begin{split}
\begin{aligned}
\pi(O(\mathbf{s}_0, t^{\prime}) | \mathbf{o})=& \int \pi\left(O\left(\mathbf{s}_0, t^{\prime}\right) | V\left(\mathbf{s}_0, t^{\prime}\right), \sigma_{\epsilon}^{2}\right) \\ & \times \pi\left(V\left(\mathbf{s}_0, t^{\prime}\right) | \boldsymbol{\theta}, \mathbf{V}, \mathbf{o}^*\right) \\
& \times \pi\left(\boldsymbol{\theta}, \mathbf{V}, \mathbf{o}^* | \mathbf{o}\right) d V\left(\mathbf{s}_0, t^{\prime}\right) d \mathbf{V} d \boldsymbol{\theta} d \mathbf{o}^*
\end{aligned}
\end{split}
\end{equation}
First, the $\theta^{(j)}$ is sampled from its marginal distribution, and $V^{(j)}$ is drawn from the posterior distribution $\pi (\theta,V,\mathbf{o}^* | \mathbf{o})$. Second, Bayesian Kriging~\cite{Bay_Krig} is applied to obtain the values of $V^{(j)}(s_0,t')$ by drawing a sample from the conditional distribution $V(s_0,t')=(O(s_1,t),\dots,O(s_n,t'))$. Third, $O^{(j)}(s_0,t')$ is drawn from the most recent $\pi\left(O\left(\mathbf{s}_0, t^{\prime}\right) | V\left(\mathbf{s}_0, t^{\prime}\right), \sigma_{\epsilon}^{2}\right)$. Fourth, summarize the samples $O^{(j)}(s_0,t')$ to { predict values at the end of the MCMC run $j \in \{1,\dots, J\}$, where $J$ is a large number of iterations. Finally, predicted values, $O(\mathbf{s}_0, t^{\prime}|\mathbf{o})$ } are transformed back to the original scale before summarizing them.

During the training, the process measurement of the model performance can be computed using the predictive model choices criteria (PMCC) (Eq.~\ref{eqn_pmcc}) as shown below, where $O\left(\mathbf{s}_{i}, t\right)_{\mathrm{pred}}$ denotes the prediction of the data $O\left(\mathbf{s}_{i}, t\right)$. The PMCC is straightforward to compute as it includes a square error loss term plus a variance term.

\begin{equation}
\label{eqn_pmcc}
PMCC= \sum_{i=1}^n \sum_{t=1}^{T}\mathbb{E}\left[O\left(\mathbf{s}_{i}, t\right)_{pred}-O\left(\mathbf{s}_{i}, t\right)\right]^{2}
+  \sum_{i=1}^n \sum_{t=1}^{T}\operatorname{Var}\left(O\left(\mathbf{s}_{i}, t\right)_{pred}\right)
\end{equation}
The PMCC is not very different from the other loss functions provided at Eq.\ref{eqn_errors}, except that it adds a variance term.

\section{Results and Analysis}
\label{sec:4}
In this section, we introduce the experimental settings in greater detail. The experiment has three components, including data prepossessing, spatial-temporal modeling and predicting, and results and analysis. 

\subsection{Data Prepossessing and Exploratory Data Analysis}

\begin{figure}[!h]
    \centering
    \includegraphics[width=3.5 in]{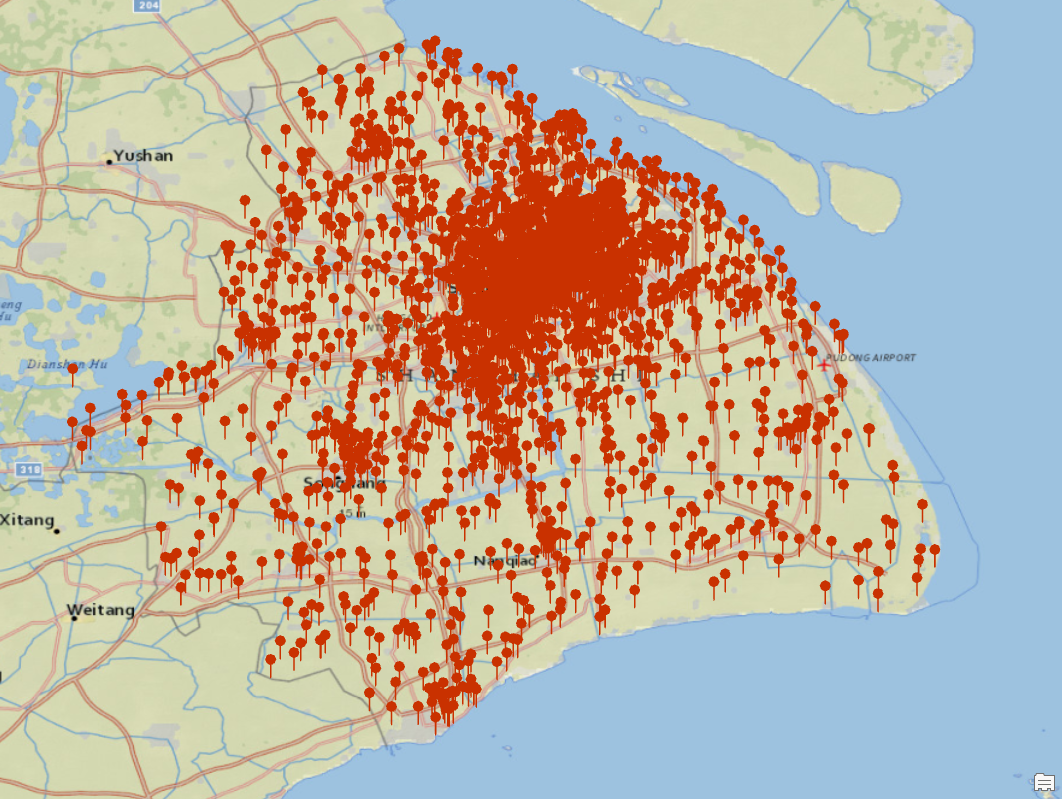}
    \caption{2\% Of The Connections In The China Telecom Dataset.}
    \label{fig:raw_data}
\end{figure}

We verify the proposed model using a real-world dataset { which is free  to download\footnote{http://sguangwang.com/TelecomDataset.html} for the research community.} Specifically, the dataset is obtained from the China Telecom company's six-month monitoring data in the Shanghai area. The dataset contains about $4.2\times10^6$ records for about 10,000 mobile users and 3,233 base stations. The dataset is considerably large, and some base stations are very close to each other, sometimes with {overlapping service areas.} It is unnecessary to predict spatial and temporal workload for each base station. Therefore, we perform some data prepossessing to clean and reduce computational costs. As we can see from Fig.\ref{fig:raw_data}, 80,000 connections (about 2\% of the total records) are messy and not suitable to show on the map, with many base stations packed in a small area; more connections are shown in the same locations if we plot {more than 2\%.} The computational cost will be considerably high if we attempt to model and predict each base station. In addition, a large number of predictions would be overlapped as the base stations are so close to each other. In other words, a workload of the same areas would be computed multiple times, and such unnecessary computation will be wasteful. Furthermore, some base stations have only a few historical connections that insufficient to train a model, which will lead to poor a performance model.

\begin{figure}[!h]
  \centering
  \begin{minipage}[b]{0.4\textwidth}
    \includegraphics[width=2.6in]{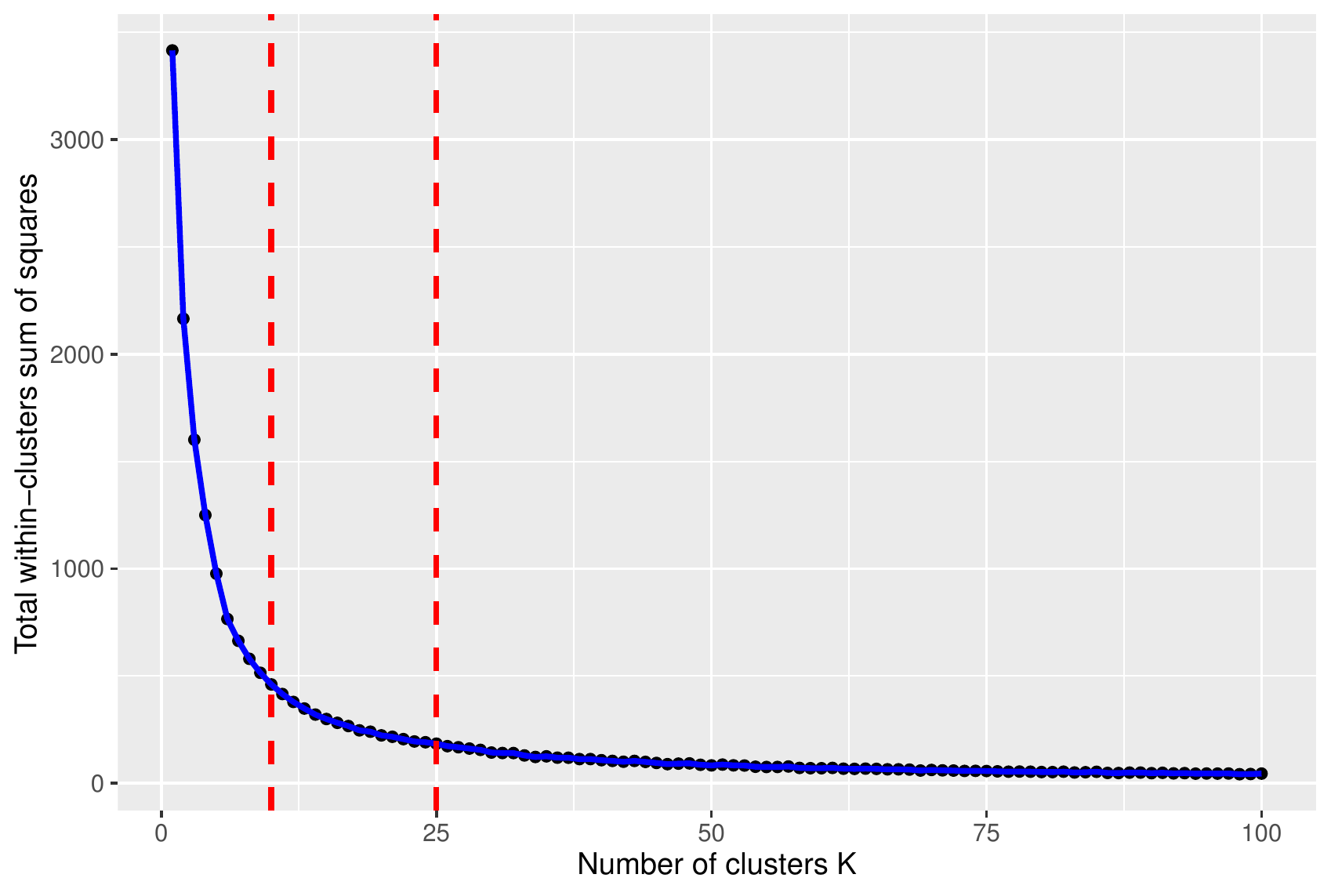}
    \caption{Elbow Method for Selecting K.}
    \label{fig:elbow}
  \end{minipage}
  \hfill
  \begin{minipage}[b]{0.4\textwidth}
    \includegraphics[width=2.2in]{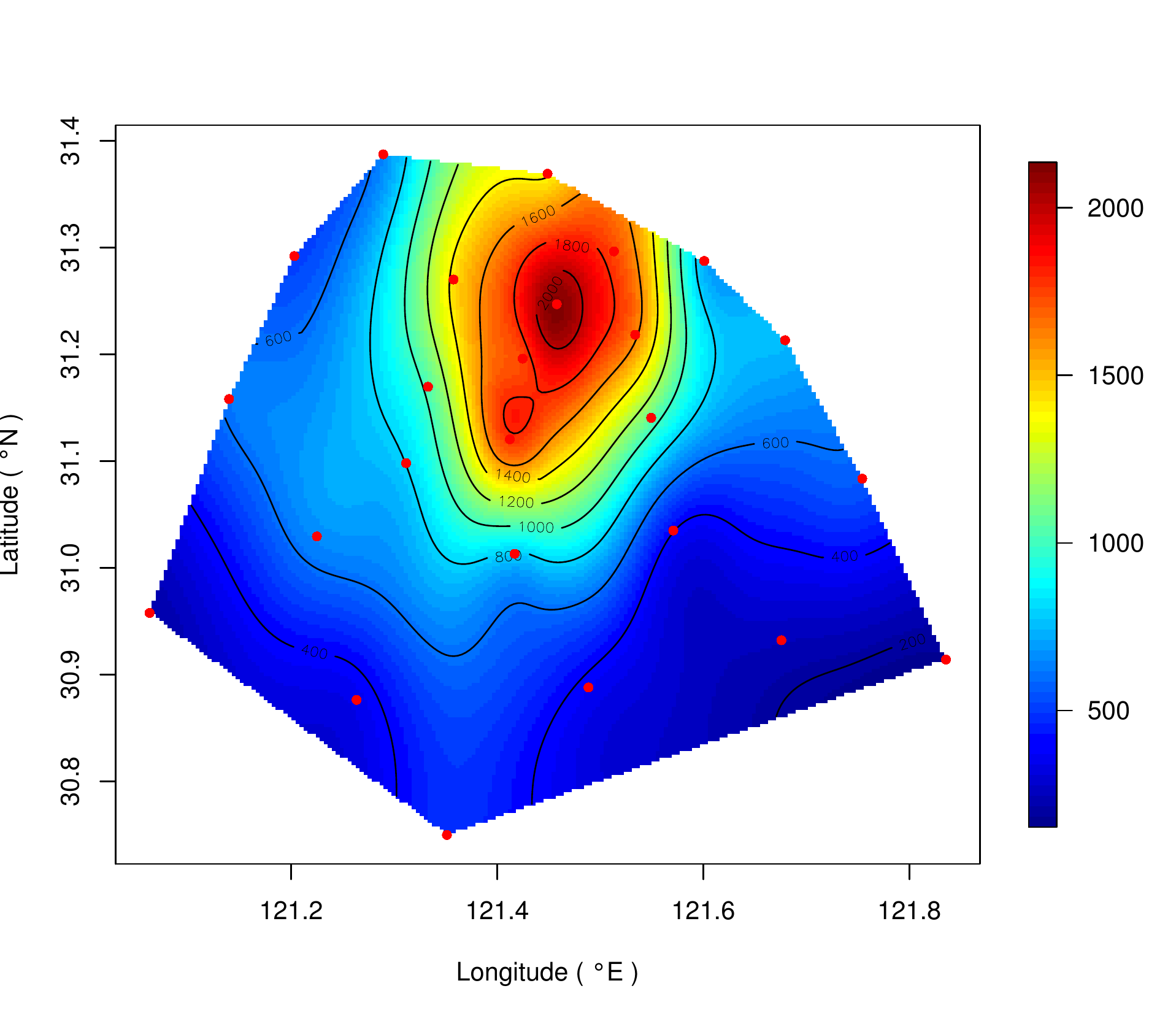}
    \caption{Distribution of Workload.}
    \label{fig:clustered}
  \end{minipage}
\end{figure}

To mitigate these issues, we cluster the data with K-Means, which is a straightforward data clustering method. The number of the clusters is chosen based on the Elbow Method, as shown in Fig.~\ref{fig:elbow}. The elbow range is from 10 to 25, and that is the reason we choose the $ K $ from that range. There is no strictly optimal value for the number of clusters. The choice of $K$ value in a reasonable range is an art rather than a science. In this study, we choose 25 as the $ K $ value because we want a relatively high resolution (i.e., the average distance to the centroid points is 200 meters). 


So, the data is clustered into 25 groups with K-Means based on the base station locations. Further, we aggregated the workload with respect to the clusters, and the contour plot is shown in Fig.~\ref{fig:clustered}. The color bar represents the clustered workload could from the blue for low values to red for high values. The workload is defined with the sum of the total time of users connected to the stations, and we can notice that the highest workload is shown in the north-middle area in this dataset. 

\begin{figure}[!h]
  \centering
  \begin{minipage}[b]{0.4\textwidth}
   \includegraphics[width=3 in]{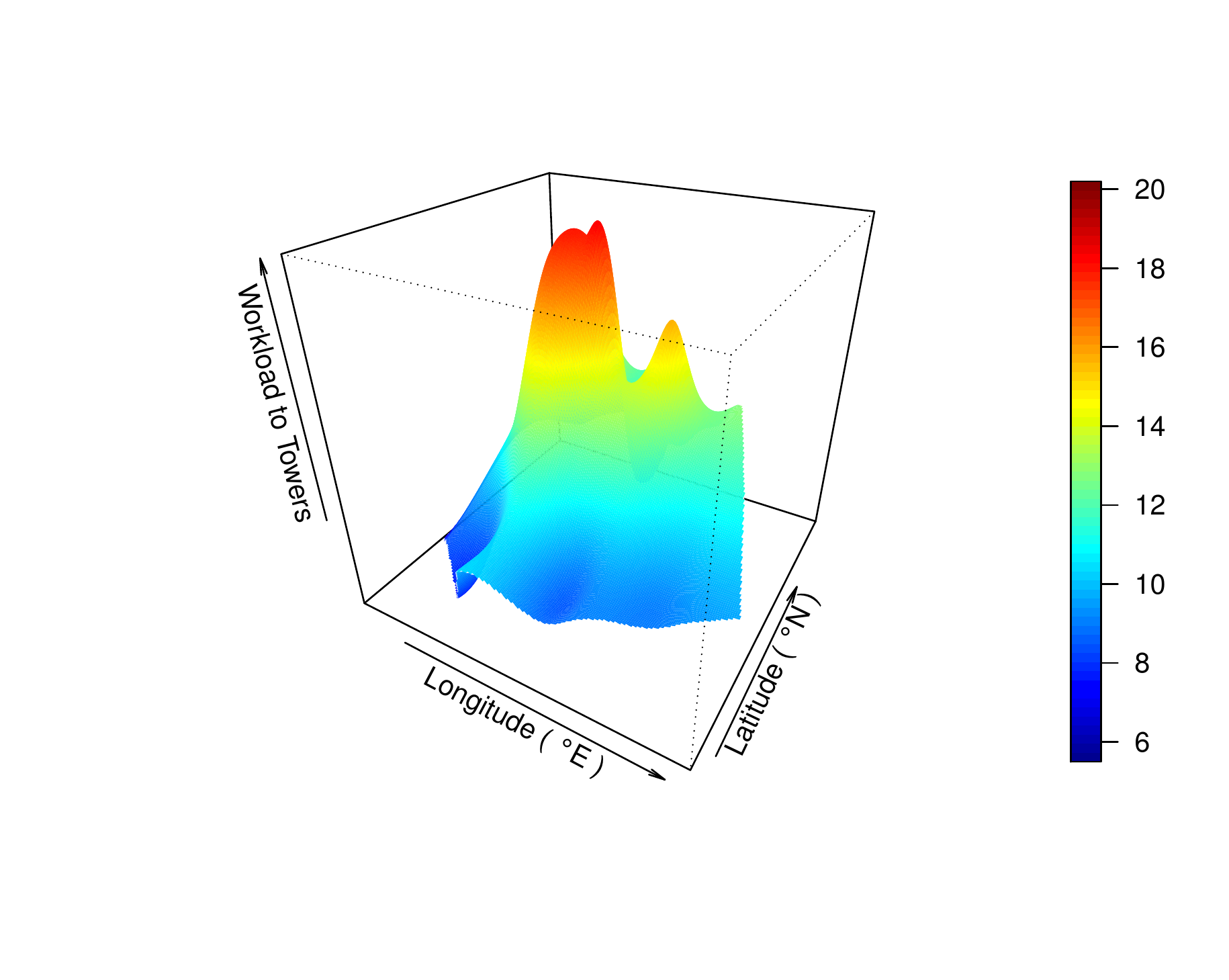}
    \caption{Workload to Number of Towers.}
    \label{fig:time_to_base_3d}
  \end{minipage}
  \hfill
  \begin{minipage}[b]{0.4\textwidth}
      \includegraphics[width=2.1in]{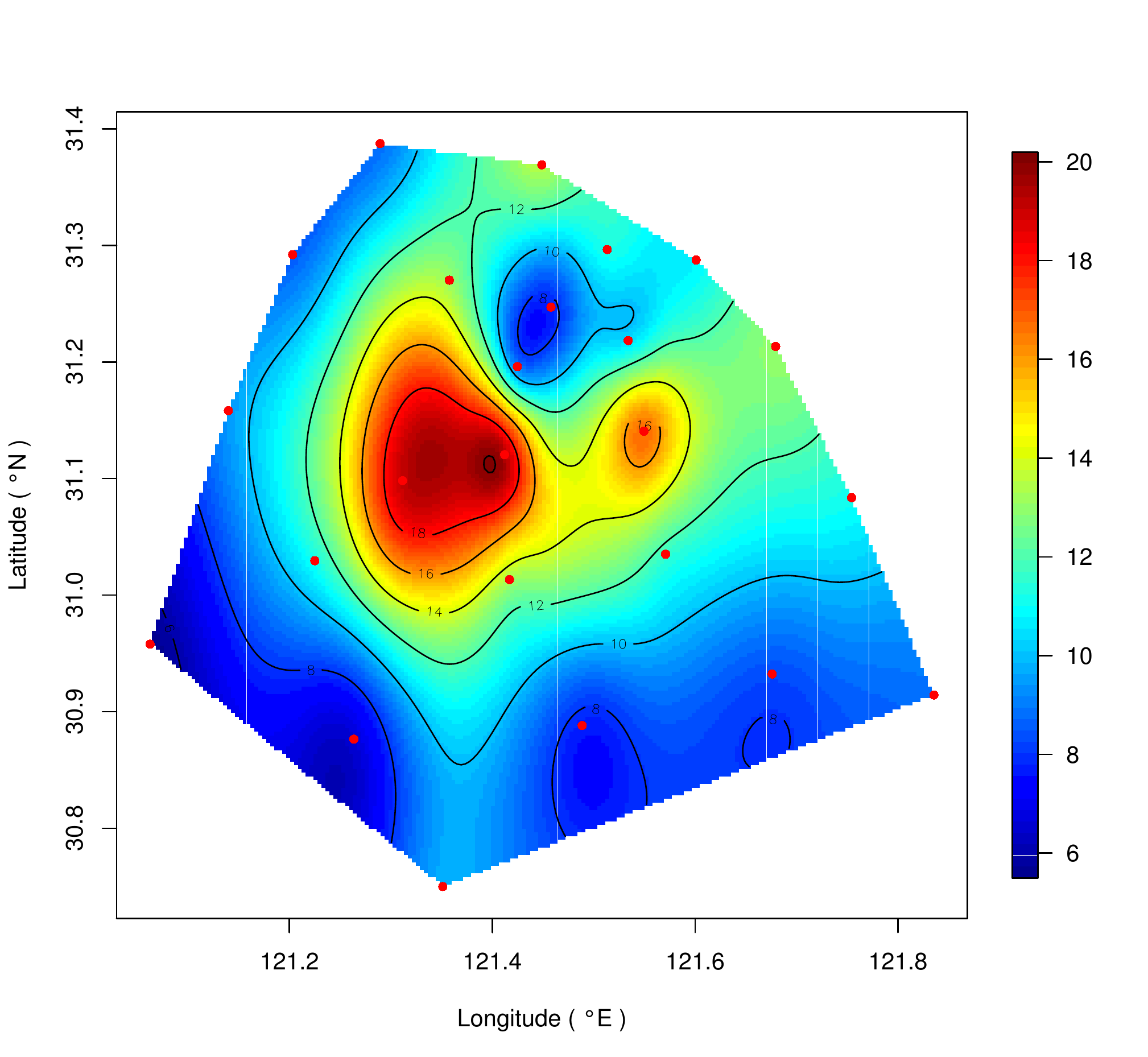}
    \caption{Contour of Fig.\ref{fig:time_to_base_3d}.}
    \label{fig:time_to_base}
  \end{minipage}
\end{figure}

Although the aggregated data shows the spatial distribution of resource demands, it is challenging to relate it to workloads of base stations. To illustrate the average workloads of towers in the areas, we compute the ratio of average workloads for each base station, as shown in Fig.~\ref{fig:time_to_base_3d} and its contour plot is shown in Fig.~\ref{fig:time_to_base}. 


We can notice that the average workload distribution, shown in Fig.~\ref{fig:time_to_base}, is quite different from the total workload shown in Fig.~\ref{fig:clustered}, which indicates that the current base station locations are not optimally distributed. Admittedly, a possible reason for the disproportional distribution of the average workload with respect to the spatial locations may be the rapid changing of resource demands. Some locations may have raised more skyscrapers in some areas of the city and changed the distribution of users and IoTs. In such circumstances, it is more meaningful to the Telecom company to predict the average workload distributions so that they can better plan future tower deployment and distribute resources more efficiently.

In order to predict the average workload distribution over space and time, we also need to extract other critical features such as the number of unique users and towers of the divided spatial areas. Specifically, the features that need extraction may include time slots, spatial coordinates, number of users and towers, total workload (connection time), and these features are transformed to logarithmic scales to reduce the noise introduced by the outliers.


\subsection{Modeling and Predicting}

With the above data prepossessing and feature engineering, we can easily split the preprocessed data into two datasets; one dataset is to train the model, and the other one is to verify the model. Specifically, we randomly chose ten location sites and selected the corresponding data as the testing set and the rest of the data as a training set. During the training of the model, we ran 5000 iterations with 1,000 \textit{burns} (i.e., it excludes the first 1,000 iterations to discard the low probability samples generated at the beginning of the training) as the regular process-based model. { The distributions of the parameters are shown below in Table.~\ref{tabke:para}.}




\begin{table}
\caption{Parameter Distributions of Auto-Regressive model}
\begin{center}
\label{tabke:para}
\begin{tabular}{@{}l S[table-format=5.4] S[table-format=5.4]  S[table-format=5.4]  S[table-format=5.4] S[table-format=5.4] @{}}
    \toprule
    \multicolumn{1}{c}{Parameters} & \multicolumn{1}{c}{Mean} & \multicolumn{1}{c}{Median} & \multicolumn{1}{c}{SD} & \multicolumn{1}{c}{Low2.5p} & \multicolumn{1}{c}{Up97.5p} \\  
    \midrule
 $\beta_0 (Intercept)$ & -0.1254 & -0.1256 & 0.0228 & -0.1706 & -0.0808 \\
$\beta_1$    & -0.0329 & -0.0328 & 0.0103 & -0.0538 & -0.0122 \\
$\beta_2 $ &  0.1667 & 0.1666 & 0.0076 & 0.1519 & 0.1815 \\
$\beta_3$   & -1.2078 & -1.2078 & 0.0039 & -1.2153 & -1.2001 \\
$\beta_4$   & 1.0316 & 1.0315 & 0.0104 &  1.0115 & 1.0524 \\

$\rho$  &  0.0233 & 0.0232 & 0.0025 & 0.0185 & 0.0281 \\
$\sigma_\epsilon^2$     &  0.0061 &  0.0061 &  0.0003 &  0.0056 &  0.0066 \\
$\sigma_\eta^2$      & 0.0107 & 0.0108 & 0.0011 & 0.0085  & 0.0126 \\
$\phi$         &  0.0019 & 0.0017 & 0.0008 & 0.0010  & 0.0040 \\
$\nu$         &  0.1500 &  0.1500 & 0.0000 &  0.1500 &  0.1500   \\
    \bottomrule
  \end{tabular}
 \end{center}
\end{table}

\begin{figure}[!h]
    \centering
    \includegraphics[width=3.2in,height=2.5in]{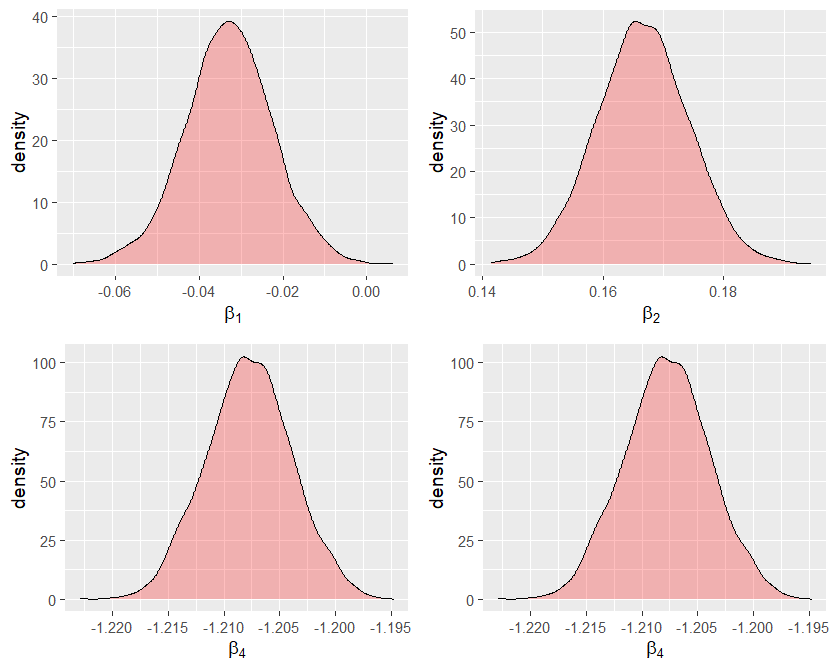}
    \caption{Distributions of Coefficients.}
    \label{fig:dist_co}
\end{figure}

\begin{figure}[!h]
    \centering
    \includegraphics[width=3.2in,height=2.5in]{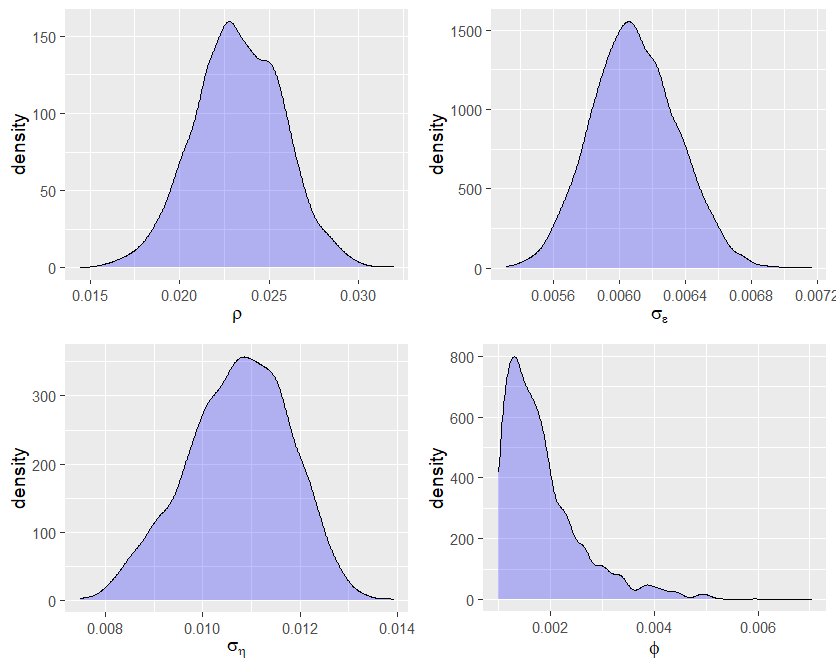}
    \caption{Distributions of Parameters.}
    \label{fig:dist_effects}
\end{figure}

\begin{table}[!h]\
\begin{center}
\caption{Performance of Trained Models}
\label{tabke:perf}
\begin{tabular}{p{6cm}|p{2.5cm}|p{2.5cm}}
Loss & Auto-Regressive & Gaussian Process  \\
\hline
Predictive Model Choice Criteria (PMCC) & 510.04 & 355.68  \\
Accept Rate, $\phi$ & 32.06\%  & 54.9\%   \\
Mean Squared Error (MSE) & 12.6706 & 12.5202  \\
Mean Absolute Error (MAE) & 2.8269 & 2.875  \\
Mean Absolute Percentage Error (MAPE) & 28.5274 & 28.4462 \\
Relative Mean Separation (RMSEP) & 3.4450 & 3.3761  \\
Relative Mean Squared Error (RMSE) & 3.5596 & 3.5384  
\end{tabular}
\end{center}
\end{table}

{ The proposed model is based on a hierarchical Bayesian model, the parameters with distinct distributions respectively (shown in Table.~\ref{tabke:para}), include the intercept, coefficients of four features ($\beta_1$: the number of connections, $\beta_2$: users, $\beta_3$: sites, and $\beta_4$: the sum of connection time), control parameter $\nu$, and effect parameters containing $\rho$, $\sigma_\epsilon^2$, $\sigma_\eta^2$, and $\phi$ .} In the posterior distribution (Eq.~\ref{eqn_ar}), the features and coefficients are denoted as $\mathbf{X}_t$ and $\boldsymbol{\beta}$, respectively. Further, we can verify the model by measuring its performance matrices listed in Table.~\ref{tabke:perf} and the acceptance rate for $\phi$ is 32\%, which falls between 15\% to 40\% recommended values shown at~\cite{Gelman2003}. The performance of the two models is similar in most of the loss matrices, and the GP model is slightly better than the AR model. 

Additionally, the distributions of the parameters make the model easier to interpret than deep learning models. Similar to the simple linear regression, the coefficients indicate the relationship between the input factors and the predicted results. For example, we can quickly notice that the log scale number of users $\beta_2$ has a positive relationship with average workloads output. The spatial and temporal correlation can also show the distributions of $\nu$ and $\rho$. Besides, the spatial correlation decay is given by the distribution of $\phi$; further, the pure and spatial-temporal variances are given by the distributions of $\sigma_\epsilon^2$ and $\sigma_\eta^2$. Fig.~\ref{fig:dist_co} shows the distributions of the regression coefficients that describe the correlations between the workload and considered factors. Similarly, the distributions of the other effects and parameters are shown in Fig.~\ref{fig:dist_effects}.

\begin{figure}[!h]
    \centering
    \includegraphics[width=5in]{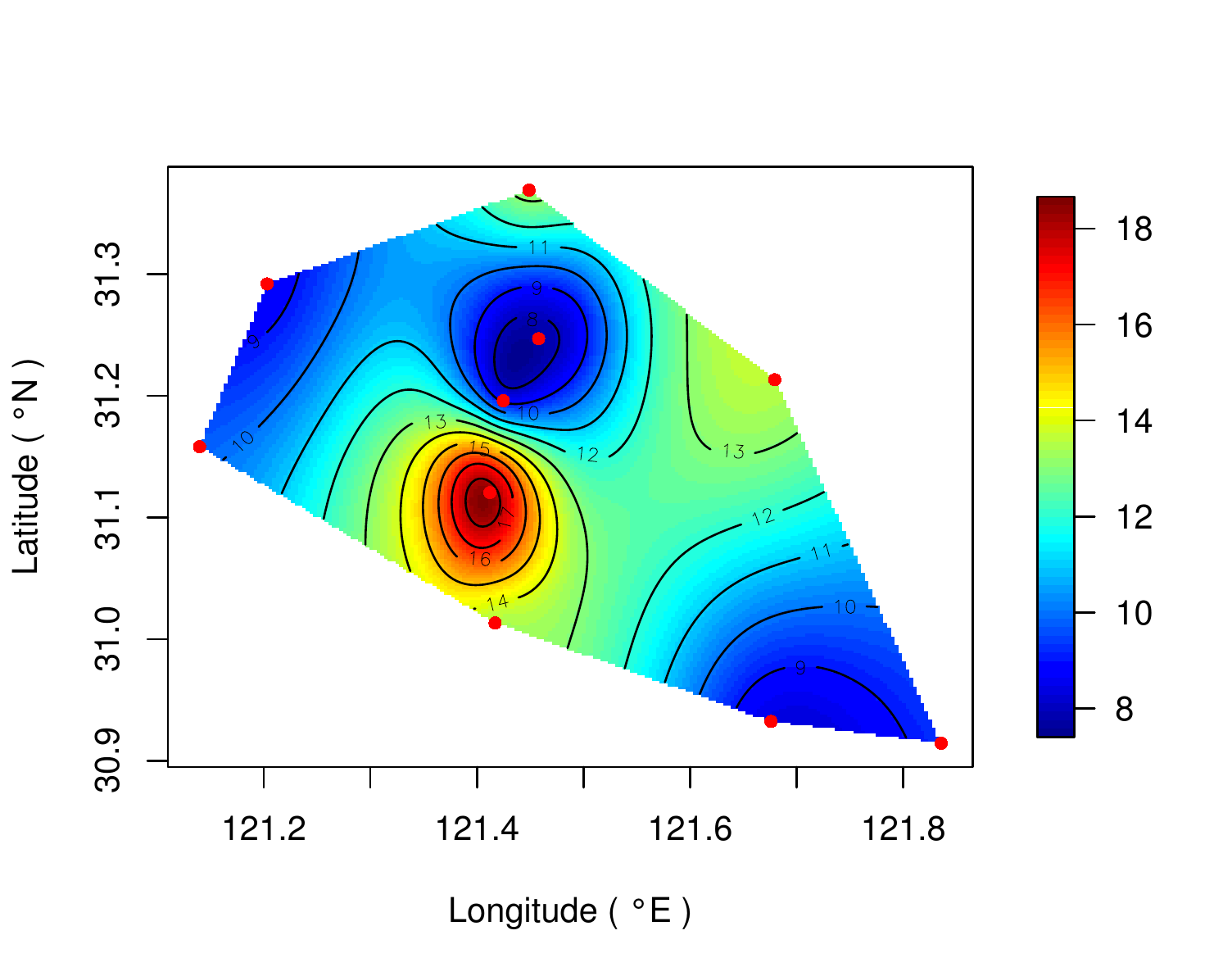}
    \caption{Spatial Prediction Mean ($\mu$).}
    \label{fig:spat_pred}
\end{figure}

{Once the model is trained, we can input unobserved locations into the model to obtain predictions, as shown in Fig.~\ref{fig:spat_pred}.} The predicted results show some areas have a higher average workload than others, and some are relatively low; therefore, the Telecom company can add more towers at the high workload areas such as the south-west area in this prediction. They can also move some of the towers from low workload areas to the high workload areas to balance the workload.

\begin{figure}[!h]
    \centering
    \includegraphics[width=3.5in]{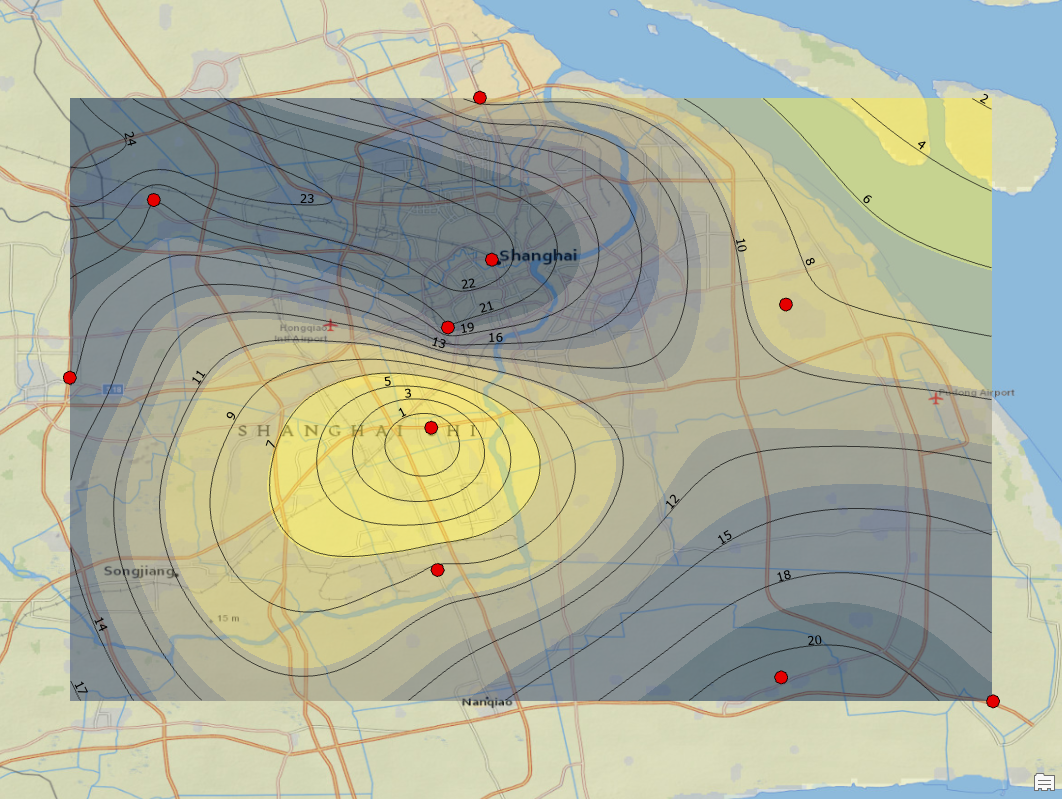}
    \caption{Prediction Mean ($\mu$) of Workload on Map.}
    \label{fig:spat_pred_map}
\end{figure}

In real-world applications, we can plot the predicted results over the real map, as shown in Fig.\ref{fig:spat_pred_map}. Note that the contour lines are slightly different from the ones shown in Fig.~\ref{fig:spat_pred} because different tools adopted different interpolation kernel functions to generate them.

\begin{figure}[!h]
  \centering
  \begin{minipage}[b]{0.4\textwidth}
    \includegraphics[width=2.3in]{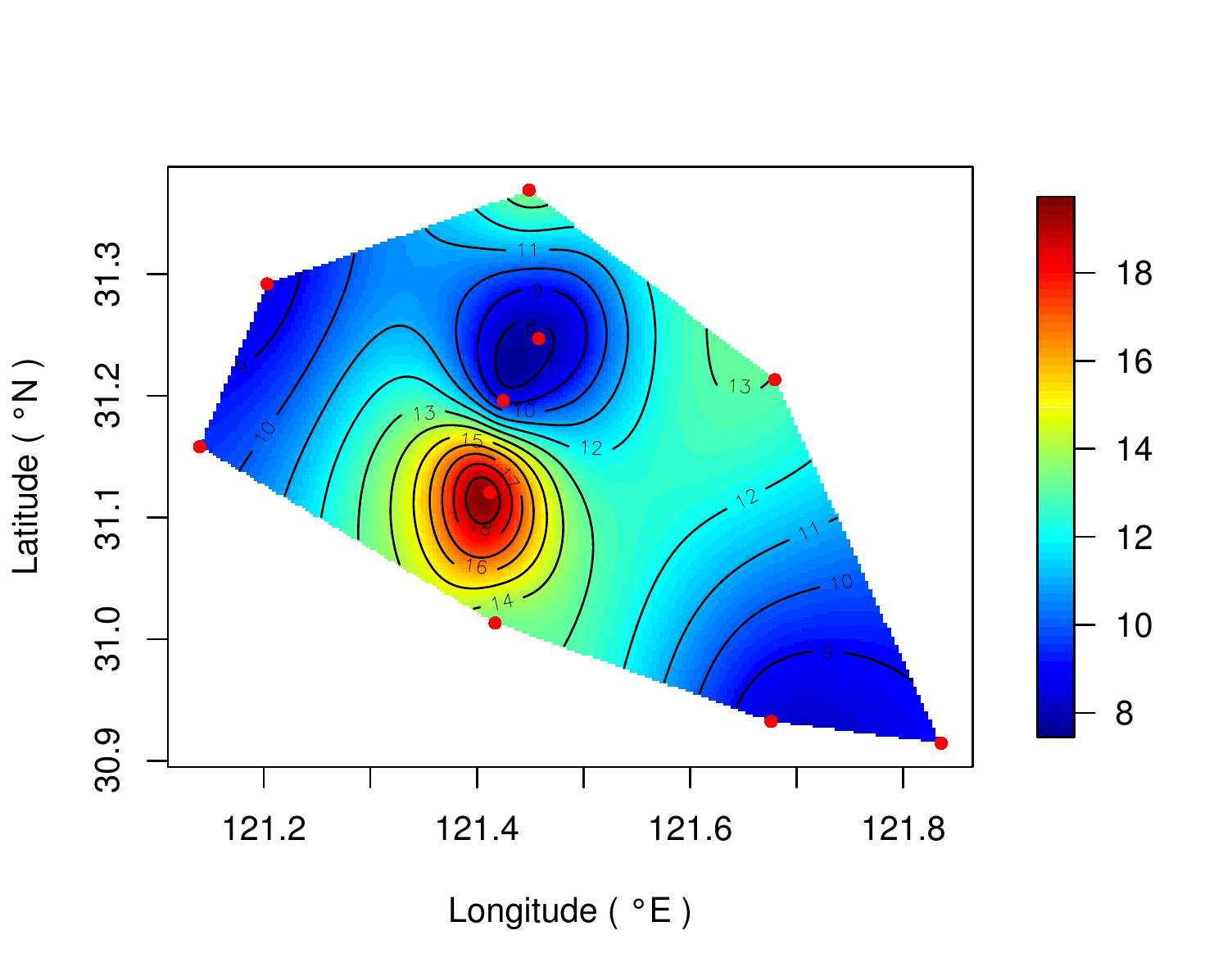}
    \caption{True label $y$ .}
    \label{fig:true_label}
  \end{minipage}
  \hfill
  \begin{minipage}[b]{0.4\textwidth}
    \includegraphics[width=2.3in]{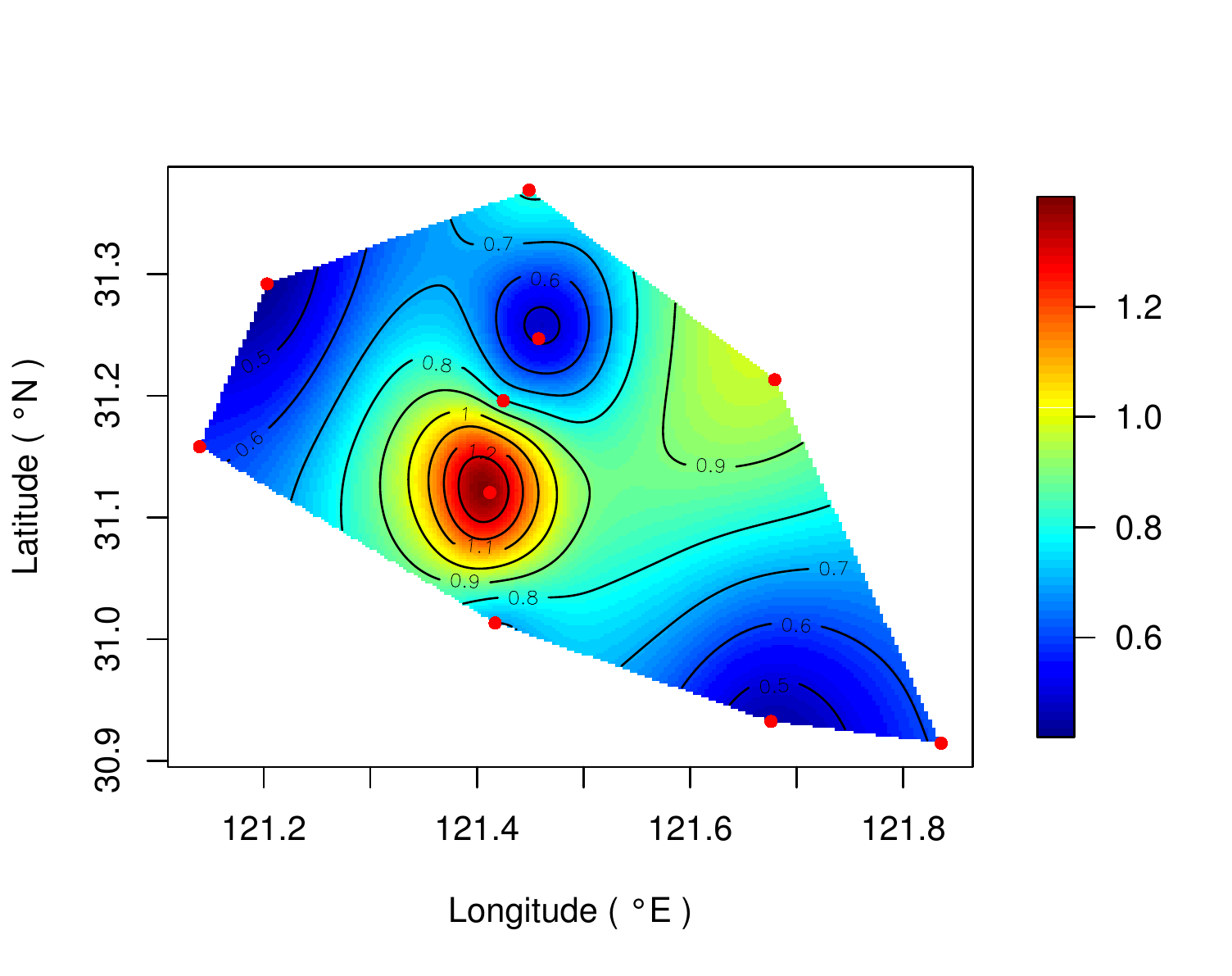}
    \caption{$L=|y-\mu|$.}
    \label{fig:loss_mean}
  \end{minipage}
  \newline
   \begin{minipage}[b]{0.4\textwidth}
    \includegraphics[width=2.3in]{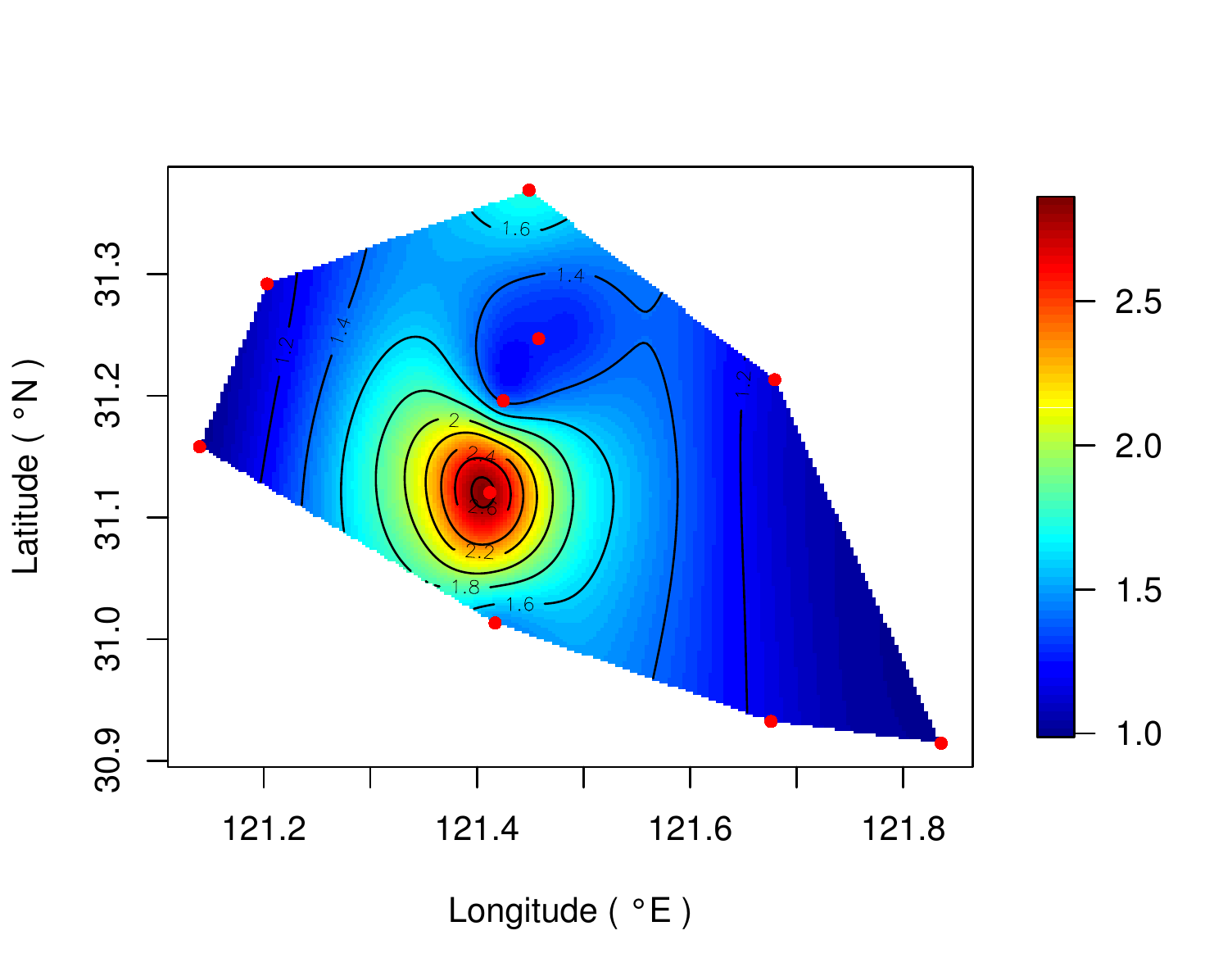}
    \caption{$L=|y-(\mu-2 \sqrt{\sigma})|$.}
    \label{fig:loss_minus_2sd}
  \end{minipage}
  \hfill
  \begin{minipage}[b]{0.4\textwidth}
    \includegraphics[width=2.3in]{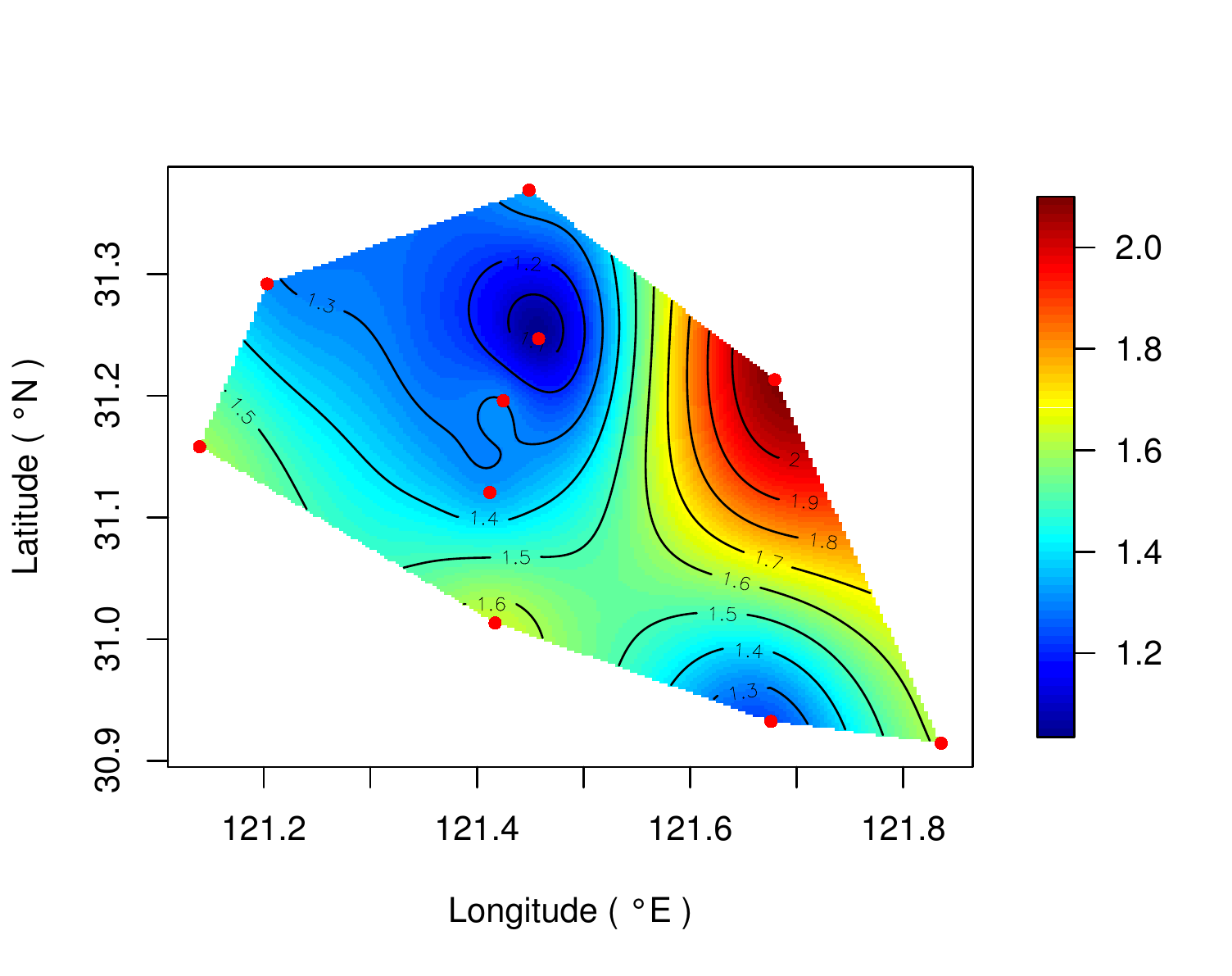}
    \caption{$L=|y-(\mu+2 \sqrt{\sigma})|$.}
    \label{fig:loss_plus_2sd}
  \end{minipage}
\end{figure}

For analyzing performance, we can compare the true label data with the prediction distribution. As the posterior and predicted values are assumed to be normally distributed, we can compute the absolute value of the loss by computing the differences of the true values (Fig.\ref{fig:true_label}) with the predicted mean $\mu$ (Fig.\ref{fig:loss_mean}), mean minus two standard deviations (Fig.\ref{fig:loss_minus_2sd}), and mean plus two standard deviations (Fig.\ref{fig:loss_plus_2sd}). Based on the empirical rule and maps shown on the figures, we can see the distributions of errors over the regions, and 95\% of the predicted values fall into this range.

\begin{figure}[!h]
    \centering
    \captionsetup{justification=centering}
    \includegraphics[width=6in]{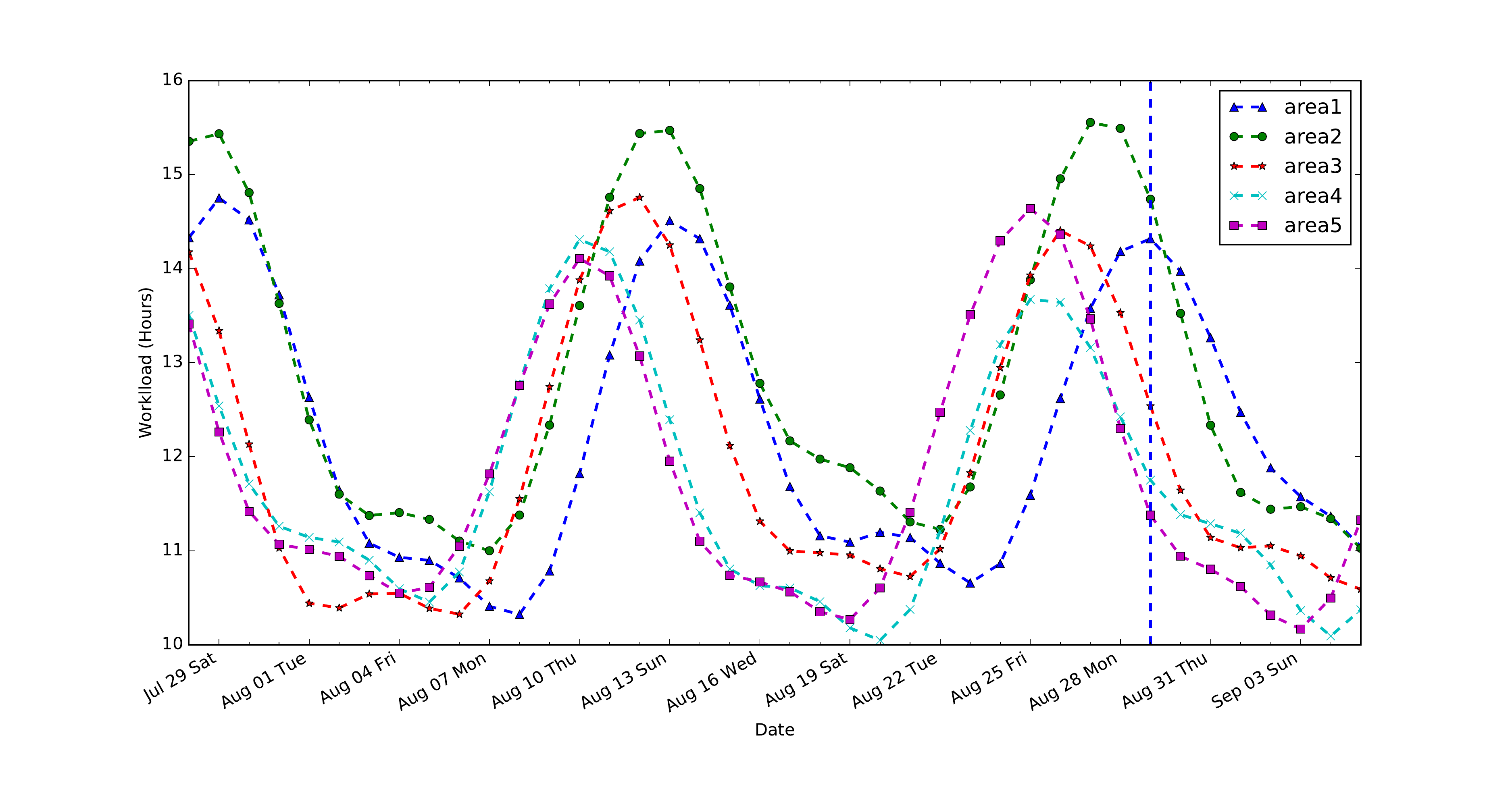}
    \caption{Temporal Prediction of Workload Values.\\ (Note: the left side of vertical dashed line are observed and the right are predictions.)}
    \label{fig:tem_pred}
\end{figure}

{Similarly, the model can predict future workload for areas, Fig.~\ref{fig:tem_pred} shows an example of future predictions for five areas with 50 days of historical demand change over time and 7 days of predicted values shown on the right side of the vertical dashed line. The original observed and predicted values are considerably noisy; therefore, the values are smoothed before plotting in Fig.~\ref{fig:tem_pred}. Specifically, the values are smoothed with a Gaussian function with a standard deviation of 2 before the plot. As we can see from Fig.~\ref{fig:tem_pred}, there is a periodic pattern of resource demands in all of the locations; most of the weekends have lower resource demands than weekdays.} Temporal learning and prediction can support future resource planning and deployment. With an accurate prediction of the future workload distribution, the Telecom company can prepare the demanded resources in advance and be able to improve the user experience within a reasonable cost.

We believe that the network vendors can deploy and plan their MEC resources efficiently using accurate spatial-temporal predictions. With reliable spatial-temporal prediction, the Telecom companies can find optimal locations for towers and deploy them according to the predicted resource demands. Additionally, they can distribute movable towers according to the predictions so that the limited resources are more efficiently used. Further, the user association system also can control the task offloading based on the prediction. For example, if a user can reach multiple towers,  { we can associate that user to the tower with the a minimum work in the future so that users can be responded to with less latency.  }

\section{Simulation}
\label{sec:5}
To demonstrate the application in mobile edge computing, we design a reasonable simulation to show the advantages of the proposed model. We assume the network provider has a limited budget to afford computational resources and deploy them to the locations. 

First, the tasks are generated by the simulator, {with the number} of tasks the same as the number of connections. Each task contains the number of required CPU cycles to process the task and its maximum tolerance. We can reasonably consider a task is expired, and the MEC servers fail to respond to the user if the process task time in the MEC server is longer than the maximum tolerance time. We generate the tasks with required CPU cycles and the maximum tolerance using the normal distributions with $c \sim \mathcal{N}(\mu_c,\sigma_c^2)$ and $\delta \sim \mathcal{N}(\mu_\delta,\sigma_\delta^2)$, respectively. The details of the parameters and their value settings are summarized in Table.\ref{tabke:sim}.
\begin{table}[!h]\
\begin{center}
\caption{Simulation Settings}
\label{tabke:sim}
\begin{tabular}{p{6cm}|p{2cm}|p{2cm}}
Description & Parameters & Values  \\
\hline
Mean of Required CPU Cycles of Tasks  & $\mu_c$ & 10 \\
Variance of Required CPU Cycles of Tasks & $\sigma_c^2$ &3 \\
Mean of Maximum Tolerant of Tasks & $\mu_\delta$ & $\mu_c /2$ \\
Variance of Maximum Tolerant of Tasks & $\sigma_\delta^2$ & 2 \\
Total Required CPU Cycles  & $\mathcal{C}$ & 1000 \\
Number of Tasks Per Time Slot & $u$ & 107 to 4501 \\
Number of the Locations & $l $ & 10 \\
Mean of Movable Tower Set On Time & $\mu_\gamma$ & 0.7 \\
Variance of Movable Tower Set On Time  & $\sigma_\gamma^2$ & 0.3 \\
\end{tabular}
\end{center}
\end{table}

Second, we assume that tasks that arrive at the same time in the same area will share the resources equally. In other words, each task shares CPU resources $c_{i,j}/u$, where and $c_{i,j}$ is the total CPU cycles distributed to location $i$ at time slot $j$, and $u$ is the number of the connections in location $i$ at time slot $j$

Third, the CPU resources are equally distributed to new locations without predicting guidance. The total CPU cycles $c_{i,j} = \mathcal{C}/l$, where $\mathcal{C}$ is the total CPU cycles that the network provider wants to distribute to all the locations, and $l$ is the number of locations. On the other hand, with prediction model guidance, they can distribute the resources weighted by the predicted results. In this simulation, we also consider the resources to be dynamically distributed as we assume there are movable towers that can be dynamically deployed. However, the movable tower resources may have some latency, and we consider this factor in the simulation. Therefore, each task arrived $i$ location at $j$ time slot can share CPU cycles:
\begin{equation}
\label{eqn_sim}
c_t = \frac{\mathcal{C}}{l} \times \gamma \times \frac{v_{i,j}}{\sum_{i,j} v_{i,j}}
\end{equation}
where, $\gamma \sim \mathcal{N} (\mu_\gamma, \sigma_\gamma^2)$ is the movable resource deploy delay factor, and $v_{i,j}$ is the prediction value for location $i$ and time slot $j$. 

\begin{figure}[h]
    \centering
    \includegraphics[width=5in]{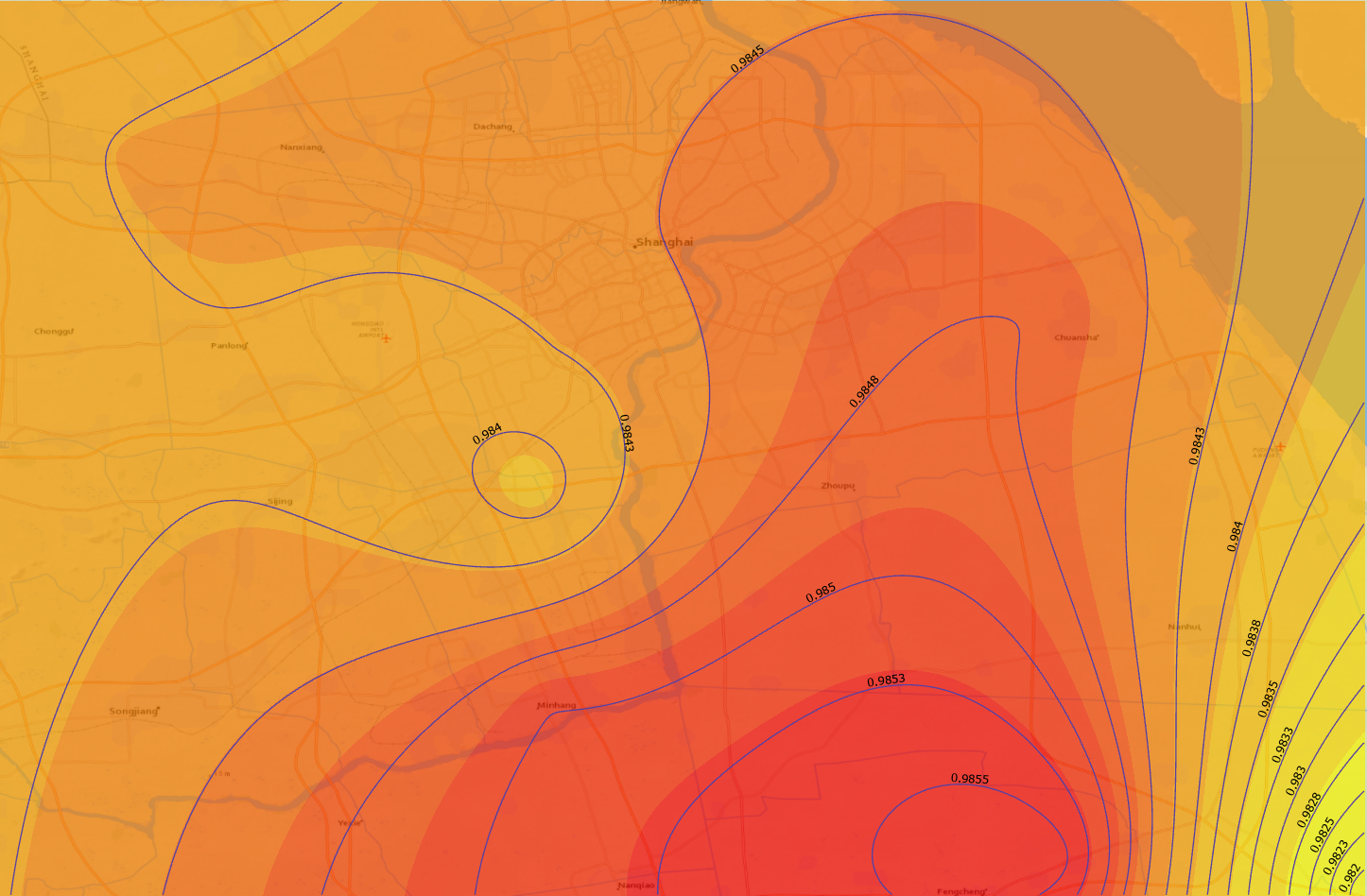}
    \caption{Completed Tasks Rate (98\%) with Prediction.}
    \label{fig:sim_opt}
\end{figure}

\begin{figure}[h]
    \centering
    \includegraphics[width=5in]{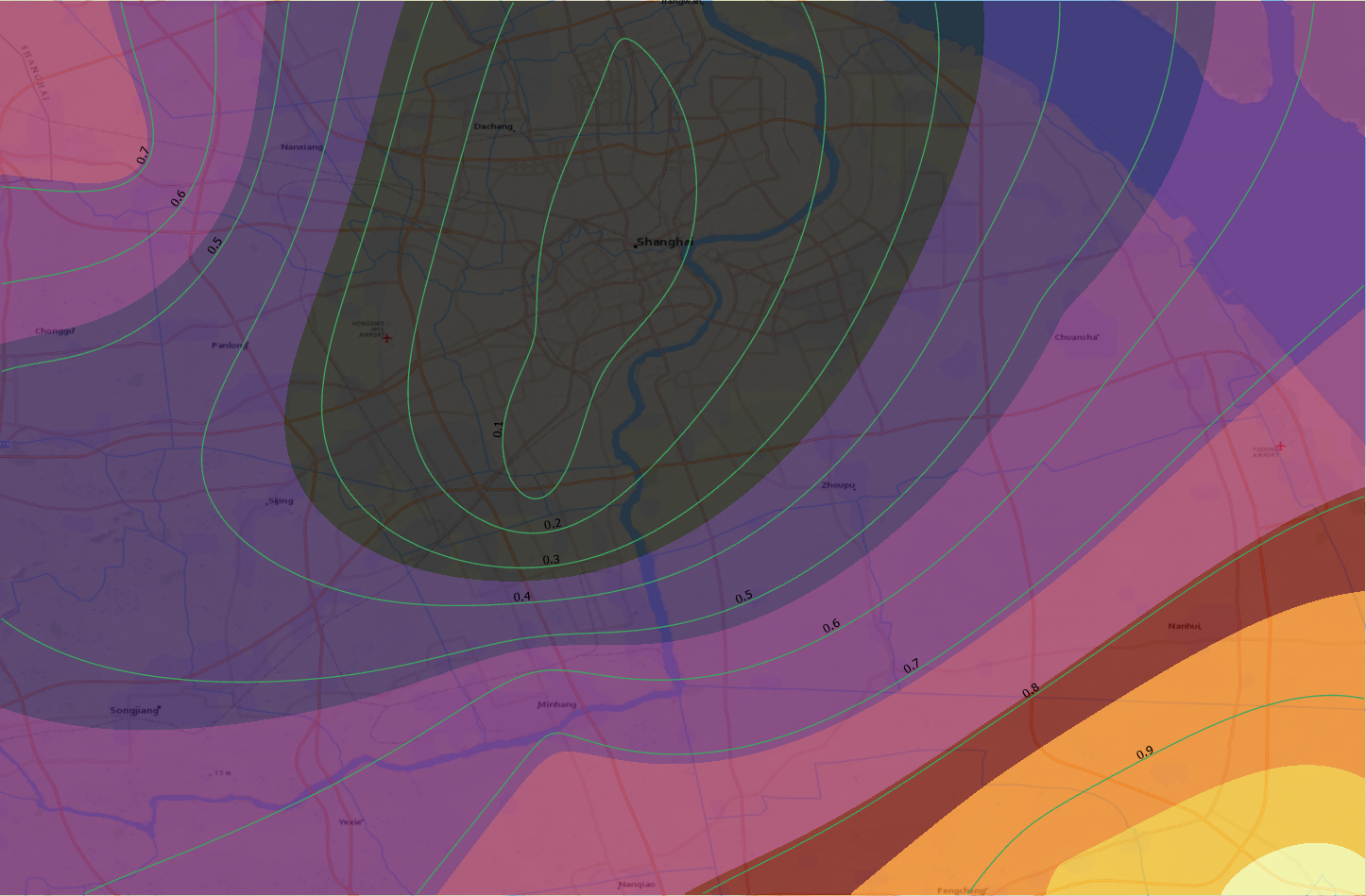}
    \caption{Completed Tasks Rate (20\%) without Prediction.}
    \label{fig:sim_non}
\end{figure}

Finally, we run the simulation 500 times and average the results because there are some random variables in the simulator. The results show the resources are more efficiently exploited when they are distributed according to predicted workload than equally distributed around the locations. As shown in Fig.~\ref{fig:sim_opt} and Fig.~\ref{fig:sim_non}, 98.41\% of offloaded tasks are successfully processed and responded to the user with a prediction model, whereas only 20.34\%
of tasks are processed before they are expired if the servers are equally distributed.

\section{Future work}
\label{sec:6}
Bayesian-based models require less training to fit the model and are relatively more straightforward than advanced models such as machine learning and deep learning. In addition, as we can see from Table.~\ref{tabke:para}, the distributions of the parameters are known to the network vendors, which makes the results more interpretable than other methods like deep learning. With the distributions of the control and coefficient parameters, the network companies know the correlations of the results and factors to prepare and adjust the resource distribution for future MEC server deployment. However, the Bayesian models have certain disadvantages, including failure to capture complex and noisy data and relying heavily on the hand setting distributions of priors. Overall, the method can provide essential data to support MEC server deployment.

In future work, we can develop algorithms for tuning and to find the optimal values to initialize the distributions of priors for the Bayesian model. Furthermore, we can leverage more geospatial data by cooperating with network companies. The data includes demographic data, transportation, building data, and other related data to increase the dimensions of the data; the more dimensions, the less uncertainty, and the model can predict more accurate results. Additionally, {network providers can also leverage the predictions for determining caching~\cite{Ale2019} demand for MECs.} Advanced task offloading models can be integrated with the proposed model to optimize the computational resources further ~\cite{Gao2020}. { Moreover,} the optimization of cooperation among the MEC servers~\cite{Chen2019} could be considered when the network providers deploy the servers. { Finally, we will investigate security issues such as the privacy data leaky~\cite{wei2020framework} and data poisoning attach~\cite{tolpegin2020data}.}

\section{conclusion}
\label{sec:7}
In this work, we adopt a hierarchical and spatial-temporal Bayesian model to model and predict MEC resource demands to support network providers to deploy MEC servers and plan resources in smart cities. The proposed model has been trained and tested on a  real-world historical dataset. To simplify the data patterns and reduce the computational cost, we preprocess and cluster the dataset into an optimal number of clusters; further, we extracted important features {in the dataset and turned them into logarithmic scales to reduce noise.} And then, the model is trained and tested with the datasets that have been preprocessed. The predicted results contain both spatial and temporal results, which means the model can predict the workload distributions for unobserved locations and future workload distributions of the observed and unobserved locations. The predicted results could provide essential information to the network provider to deploy resources and use them efficiently. Moreover, the model has plausible interpretability because the distributions of the parameters are known to the Telecom companies. Finally, we designed a simulation to show a possible application; the simulation shows {the deployed resources according to predicted results can process more offloaded tasks than using resources equally. Such} efficient use of network resources is crucial for efficient smart cities.

%
\nocite{*}
\bibliographystyle{IEEEannot}
\bibliography{annot}

%
\appendix
\section{appendix}
{ In this section,} we present the full conditional distributions of the proposed AR model, and process of deriving GP model is very similar to AR. Therefore, the full conditional distributions of the GP model is omitted.
The full conditional distribution of $\beta$ can be obtained from the joint posterior distribution of the proposed AR model (Eq.~\ref{eqn_ar}) as $\pi(\boldsymbol{\beta} | \ldots, \mathbf{z}) \sim N(\Delta \chi, \Delta)$, where:
\begin{equation}
\label{eqn_ar_beta}
\begin{aligned}
&\Delta^{-1}= \sum_{t=1}^{T} \mathbf{X}_{t}^{\top} \Sigma_{\eta}^{-1} \mathbf{X}_{l t}+\mathbf{I}_{p} / \delta_{\beta}^{2},\\
&\chi= \sum_{t=1}^{T} \mathbf{X}_{t}^{\top} \Sigma_{\eta}^{-1}\left(\mathbf{V}_{t}-\rho \mathbf{V}_{ t}\right).
\end{aligned}
\end{equation}

The full conditional distribution of $\rho$ can be obtained as $\pi(\rho| \ldots, \mathbf{O}) \sim N(\Delta \chi, \Delta)$, where:

\begin{equation}
\label{eqn_ar_rho}
\begin{aligned}
&\Delta^{-1}= \sum_{t=1}^{T} \mathbf{V}_{t}^{\top} \Sigma_{\eta}^{-1} \mathbf{V}_{t}+\mathbf{I}_{p} / \delta_{\rho}^{2},\\
&\chi= \sum_{t=1}^{T} \mathbf{V}_{t}^{\top} \Sigma_{\eta}^{-1}\left(\mathbf{V}_{t}-\mathbf{X}_{ t} \boldsymbol{\beta}\right).
\end{aligned}
\end{equation}

For $\sigma_\epsilon^2$ and $\sigma_\eta^2$ can be sampled from the following conditional distributions, respectively:
\begin{equation}
\label{eqn_ar_rho}
\begin{aligned}
\pi\left(1 / \sigma_{\epsilon}^{2} | \ldots, \mathbf{O}\right) \sim & G\Bigl(\frac{N}{2}+a, b+\frac{1}{2}  \sum_{t=1}^{T} \left(\mathbf{O}_{t}-\mathbf{V}_{ t}\right)^{\top}\left(\mathbf{O}_{t}-\mathbf{V}_{ t}\right)\Bigr),
\end{aligned}
\end{equation}

\begin{equation}
\label{eqn_ar_rho}
\begin{aligned}
\pi\left(1 / \sigma_{\eta}^{2} | \ldots, \mathbf{O}\right) \sim & G\Bigl(\frac{N}{2}+a, b+\frac{1}{2}  \sum_{t=1}^{T} \left(\mathbf{O}_{t}-\mathbf{V}_{ t}\right)^{\top}\left(\mathbf{O}_{t}-\mathbf{V}_{ t}\right)\Bigr)\\
 \sim & G\Bigl(\frac{N}{2}+a, b+\frac{1}{2} \sum_{t=1}^{T_{}}\bigl(\mathbf{V}_t-\rho \mathbf{V}_{ t-1}\\
 & - \mathbf{X}_t \boldsymbol{\beta}\bigr)^{\top} \zeta_{\eta}^{-1} 
\left(\mathbf{V}_t-\rho \mathbf{V}_{ t-1}-\mathbf{X}_t \boldsymbol{\beta}\right)\Bigr) .
\end{aligned}
\end{equation}

Based on the joint posterior (Eq.~\ref{eqn_ar}), the full conditional distribution for $V_t$ can be derived in two cases when $1 \leq t \leq T$ and when $t=T$, such that $\pi(V_t| \ldots, \mathbf{O}) \sim N(\Delta \chi_t, \Delta_t)$, where: \\Case 1:
\begin{equation}
\label{eqn_ar_rho}
\begin{aligned}
\Delta_t^{-1}=\mathbf{I}_{n} / \sigma_{\epsilon}^{2}+\left(1+\rho^{2}\right) \Sigma_{\eta}^{-1},\\
\chi_t = \mathbf{O}_t / \sigma_{\epsilon}^{2}+\Sigma_{\eta}^{-1}\left(\rho \mathbf{V}_{t-1}+\mathbf{X}_t \boldsymbol{\beta} +  \rho\left(\mathbf{V}_{t+1}-\mathbf{X}_{t+1} \boldsymbol{\beta}\right)\right).
\end{aligned}
\end{equation}
Case 2:
\begin{equation}
\label{eqn_ar_rho}
\begin{array}{c}
\Delta_{t}^{-1}=\mathbf{I}_{n} / \sigma_{\epsilon}^{2}+\Sigma_{\eta}^{-1} ,\\
\chi_{t}=\mathbf{O}_{t} / \sigma_{\epsilon}^{2}+\Sigma_{\eta}^{-1}\left(\rho \mathbf{V}_{t-1}+\mathbf{X}_{t} \boldsymbol{\beta}\right).
\end{array}
\end{equation}

The full conditional distribution for $V_0$ is $N(\Delta \chi, \Delta)$, where:
\begin{equation}
\label{eqn_ar_rho}
\begin{aligned}
&\Delta^{-1}=\rho^{2} \Sigma_{\eta}^{-1}+\frac{1}{\sigma^{2}} S_{0}^{-1},\\
&\chi=\rho\left(\mathbf{V}_{1}-\mathbf{X}_{1} \boldsymbol{\beta}\right)^{\top} \Sigma_{\eta}^{-1}+\frac{1}{\sigma^{2}} \boldsymbol{\mu}^{\top} \zeta_{0}^{-1}.
\end{aligned}
\end{equation}

The full conditional distribution for $\mu$ can be obtained from $N(\Delta \chi, \Delta)$, where:
\begin{equation}
\label{eqn_ar_rho}
\begin{aligned}
\begin{array}{c}
\Delta^{-1}=\frac{1}{\sigma^{2}} \zeta_{0}^{-1}+\frac{1}{\sigma_{\mu}^{2}} \mathbf{I}_{n} ,\\
\chi=\frac{1}{\sigma^{2}} \zeta_{0}^{-1} \mathbf{V}_{0} .
\end{array}
\end{aligned}
\end{equation}

The full conditional distribution for $\sigma^{2}$:
\begin{equation}
\label{eqn_ar_rho}
\begin{aligned}
\pi\left(1 / \sigma^{2} | \ldots, \mathbf{O}\right) \sim G\left(\frac{n}{2}+a, b+\frac{1}{2}\left(\mathbf{V}_{0}-\boldsymbol{\mu}\right)^{\top} \zeta_{0}^{-1}\left(\mathbf{V}_{0}-\boldsymbol{\mu}\right)\right).
\end{aligned}
\end{equation}

The full conditional distribution of $\phi$ parameter given by:

\begin{equation}
\label{eqn_ar_kernel}
\begin{aligned}
\pi(\phi | \ldots, \mathbf{O}) \propto & \pi(\phi) \times\left|\mathbf{\zeta}_\eta\right|^{-T / 2} \times \exp \Bigl[-\frac{1}{2 \sigma_{\eta}^{2}} \sum_{t=1}^{T}\bigl(\mathbf{V}_{t}\\ & -\rho \mathbf{V}_{t-1} -\mathbf{X}_{t} \boldsymbol{\beta}\bigr)^{\top} \zeta_{\eta}^{-1}\left(\mathbf{V}_{t}-\rho \mathbf{V}_{t-1} -\mathbf{X}_{t} \boldsymbol{\beta}\right)\Bigr] \\
& \times\left|\mathbf{\zeta}_{0}\right|^{-r / 2} \times \exp \Bigl[-\frac{1}{2} \frac{1}{\sigma^{2}}\left(\mathbf{V}_{0}-\boldsymbol{\mu}\right)^{\top} \zeta_{0}^{-1} \\ & \left(\mathbf{V}_{0}-\boldsymbol{\mu}\right)\Bigr].
\end{aligned}
\end{equation}

\end{document}